\documentclass[times,sort&compress,3p]{elsarticle}
\journal{.}
\usepackage[labelfont=bf]{caption}
\usepackage{amsmath,amsfonts,amssymb,amsthm,booktabs,color,epsfig,graphicx,url}
\usepackage{epstopdf}
\usepackage{diagbox}
\usepackage{slashbox}

\theoremstyle{plain}
\newtheorem{theorem}{Theorem}[section]
\newtheorem{proposition}{Proposition}[section]
\newtheorem{lemma}{Lemma}[section]
\newtheorem{corollary}{Corollary}[section]
\theoremstyle{definition}

\newtheorem{remark}{Remark}[section]
\newtheorem{example}{Example}[section]
\begin{document}
\begin{frontmatter}

\title{Worst-cases of distortion riskmetrics and weighted entropy with partial information}
%[label1]
\author{Baishuai  Zuo}
\author{Chuancun Yin\corref{mycorrespondingauthor}}
\address{School of Statistics and Data Science, Qufu Normal University, Qufu, Shandong 273165, P. R. China}

\cortext[mycorrespondingauthor]{Corresponding author. Email address: \url{ccyin@qfnu.edu.cn (C. Yin)}}

\begin{abstract}
In this paper, we discuss the worst-case of distortion riskmetrics for general distributions when only partial information (mean and variance) is known. This result is applicable to general class of distortion risk measures and variability measures. Furthermore, we also consider worst-case of weighted entropy for general distributions when only partial information is available. Specifically, we provide some applications for entropies, weighted entropies and risk measures.  The commonly used entropies include Gini functional, cumulative residual entropy, tail-Gini functional, cumulative Tsallis past entropy, extended Gini coefficient and so on. The risk measures contain some premium principles and shortfalls based on entropy. The shortfalls include the Gini shortfall, extended Gini shortfall, shortfall of cumulative residual entropy and shortfall of cumulative residual Tsallis entropy with order $\alpha$.
\end{abstract}

\end{frontmatter}
\section{Introduction\label{sec:1}}
Distortion riskmetrics play an important role in the construction of premium
principles and risk aversion, and also are widely used in behavioral
 economics and risk management (see, e.g., Yaari, 1987; Denneberg, 1990;  Wang, 2000; Denuit et al., 2005; Dhaene et al., 2012; Wang et al., 2020). A lot of entropies are distortion riskmetrics obtained by choosing appropriate distortion functions, such as Gini functional (Gini), cumulative residual entropy (CRE), tail-Gini functional (TGini), cumulative Tsallis past entropy (CT), extended Gini coefficient (EGini) and variability measures based on distortion functions $h$
and the CRE ($h$-CRE) etc. The entropies are regarded as a useful tool in researching life tables, the cost of annuities, and
premium principles (see, e.g., Haberman et al., 2011; Psarrakos and Sordo, 2019; Hu and Chen, 2020; Sun et al., 2022; Yin et al., 2023; Zuo and Yin, 2023; Psarrakos et al., 2024).

Making reasonable decisions in uncertain situations usually requires
quantitative analysis and the use of model. Model decision-making is sensitive to potential model deviations and data issues.  Actuarial Association of Europe (2017) also pointed out that model decision-making cannot be ignored, but the selection of models has a certain degree of subjectivity (see, for instance, Bernard et al., 2020).

In practice, for a set of data, its model or distribution is uncertain, and its full information cannot be obtained, only partial information, such as moment information, is known. The usual approach is to compute (or estimate) the first two moments (expectation and variance) of the data, and then consider the worst-case and best-case estimates of the risk measure when the first two moments are known. This direction has attracted high attention from scholars and has been extensively covered in many literature. For examples, EI Ghaoui
et al. (2003), Chen et al. (2011) and Li et al. (2018) provided closed-form solutions for the worst-case Value-at-Risk (VaR), worst-case tail Value-at-Risk (TVaR), worst-case range Value-at-Risk (RVaR) problems when the first two moments are known, respectively. As generalization of Li et al. (2018), Zhu
and Shao (2018) studied the worst-case and best-case bounds for distortion
risk measures under additional symmetry information. Liu et al. (2020) considered the worst-case
 values for a law-invariant convex risk functional under the condition that higher-order moments are known.
 Cai et al. (2023) presented the worst-case
 distribution for any general class of distortion risk measures, and provided further explanation of such
risk-averse behaviour, which is the generalization of Liu et al. (2020).
  Bernard et al. (2020) derived RVaR bounds for unimodal distributions under partial
information.  Bernard et al. (2023) studied largest (smallest) value for the class of distortion risk measures with an absolutely continuous distortion function when mean and variance of loss distribution are known.
Recently,
Shao and Zhang (2023a, 2023b) provided
closed-form solutions of distortion risk measures in extreme cases by using
 the first two moments and the symmetry of underlying distributions.
Zhao et al. (2024) studied the best- and worst-case of a general class of distortion risk measures under the condition that partial information of
 distributions is known. They all use envelopes of the distortion function to characterize extreme-case distributions.

Our research motivation is three-fold. First, distortion riskmetrics include general of distortion risk measures and signed Choquet integrals, and have a wide range of applications, such as life tables, the cost of annuities, and
premium principles.  Second, most of the existing studies mainly focus on worst-case or best-case for
general class of distortion risk measures, such as VaR, TVaR and RVaR. There is almost no involvement in distortion riskmetrics yet, especially variability measures (or entropies), such as Gini, EGini, CRE, CT and $h$-CRE and so on. Third, in practice, the question `` for a set of data, we only know its' partial information, such as moment information, to fast estimate a risk measure'' is common. To estimate sharp upper bounds of entropies, weighted entropies and
risk measures (including some premium principles and shortfalls based on entropy), we derive worst-case of distortion riskmetrics and weighted entropy for general distributions when only mean and variance are known.

Inspired by all these work, we make the following contributions in this paper. Firstly, we consider the worst-case of distortion riskmetrics for general distributions when mean and variance are known. Secondly, we extend this result to weighted entropy for general distributions when mean and variance are available. Thirdly, we provide some applications to  entropies, weighted entropies and
risk measures (containing some premium principles and shortfalls based on entropy).

The remainder of this paper is structured as follows. Section 2 provides the quantile representations of distortion riskmetrics and  weighted entropy, and also provides some notations.  Section 3 derives the worst-cases of distortion riskmetrics and weighted entropy for general distributions under the condition that mean and variance are known.  Sections 4 and 5 give applications to entropies and weighted entropies,  respectively.  In Section 6, we provide applications to risk measures, including sharp upper bounds of premium principles and worst-cases of shortfalls based on entropy.  Numerical illustration is given in Section 7. Finally, Section 8 puts some concluding remarks.

Throughout the paper,  let $(\Omega, \mathcal{F}, P)$ be an atomless probability space. Let $L^{0}$
be the set of all random variables  on $(\Omega, \mathcal{F}, P)$.
Denote by $L
^{0}_{+}$
the set of all nonnegative random variables, and denote by $L
^{\infty}$ the set of essentially bounded
random variables. $g'$ denotes (first) right
derivative of $g$. Notation
 $$\mathcal{G}=\{ g: [0, 1] \rightarrow \mathbb{R},~ g~ \mathrm{is ~of~ bounded~ variation~ and}~g(0)=0\},$$
  and $\mathbb{I}_{A}(\cdot)$ is the indicator function of set $A$.
For $(\mu, \sigma) \in \mathbb{R} \times \mathbb{R}_{+}$, we denote by $V (\mu, \sigma)$ the set of random variables
with mean $\mu$ and variance $\sigma^{2}$. Similarly, for $(\mu_{\Psi}, \sigma_{\Psi}) \in \mathbb{R} \times \mathbb{R}_{+}$, we denote by $V (\mu_{\Psi}, \sigma_{\Psi})$ the set of random variables
with mean $\mu_{\Psi}$ and variance $\sigma_{\Psi}^{2}$.
In addition,
the left-continuous generalized inverse of $F_{V}$ is defined by
$$F_{V}^{-1}(p):=\mathrm{inf}\{x\in\mathbb{R} : F_{V} (x) \geq p\},~p\in(0,1],~\mathrm{and}~F_{V}^{-1}(0):=\sup\{x\in\mathbb{R} : F_{V} (x)=0\},$$
while it's right-continuous generalized inverse is defined by
$$F_{V}^{-1+}(p):=\mathrm{sup}\{x\in\mathbb{R} : F_{V} (x) \leq p\},~p\in(0,1],~\mathrm{and}~F_{V}^{-1+}(1):=F_{V}^{-1}(1).$$

We recall definitions of  distortion riskmetric and weighted entropy (form) as follows.

A functional $\rho_{g} : \mathcal{X} \rightarrow \mathbb{R}$, whose domain $\mathcal{X} \supset L^{\infty}$ is a law-invariant convex cone, is a distortion riskmetric if there exists $g \in \mathcal{G}$ such that
\begin{align}\label{a1}
 \rho_{g}(X)=\int_{-\infty}^{0}\left[g\left(P(X\geq x)\right)-h(1)\right]\mathrm{d}x+\int_{0}^{\infty}g\left(P(X\geq x)\right)\mathrm{d}x,
\end{align}
where $g$ is the distortion function of $\rho_{g}$.
Note that distortion riskmetrics is generalizations of signed Choquet integrals (see Wang et al. (2020b)) and general class of distortion risk measures (see Dhaene et al. (2012)).

For $g\in\mathcal{G}$ and $g(1)=0$, a random variable (r.v.) $X$, with distribution function $F_{X}(x)$, tail distribution function $\overline{F}_{X}(x)$ and weighted function $\psi(x)$: $x \in \mathbb{R}\rightarrow \psi(x)\in\mathbb{R}$, has its
 the weighted entropy (form)  as
  \begin{align}\label{a2}
 \int_{-\infty}^{+\infty}\psi(x)g(\overline{F}_{X}(x))\mathrm{d}x.
\end{align}
Note that when $\psi(x)=1$ in Eq. (\ref{a2}),  it will be the usual entropy (form) defined as
\begin{align}\label{a3}
 \int_{-\infty}^{+\infty}g(\overline{F}_{X}(x))\mathrm{d}x.
\end{align}
\section{Preliminaries\label{sec:2}}
Quantile representations of distortion riskmetrics and weighted entropy are given as follow.
\begin{lemma}\label{le.1}[Wang et al., 2020a]
Let $g\in\mathcal{G}$, and $\rho_{g}$ be a distortion riskmetric defined as in Eq. (\ref{a1}). For $X\in L^{0}$, \\
($\mathrm{i}$) if $g$ is right-continuous,
\begin{align*}
   \rho_{g}(X)&=\int_{0}^{1}F_{X}^{-1+}(1-u)\mathrm{d}g(u)=\int_{0}^{1}F_{X}^{-1+}(u)\mathrm{d}\hat{g}(u);
\end{align*}
($\mathrm{ii}$) if $g$ is left-continuous,
\begin{align*}
   \rho_{g}(X)&=\int_{0}^{1}F_{X}^{-1}(1-u)\mathrm{d}g(u)=\int_{0}^{1}F_{X}^{-1}(u)\mathrm{d}\hat{g}(u);
\end{align*}
($\mathrm{iii}$) if $g$ is continuous,
\begin{align*}
   \rho_{g}(X)&=\int_{0}^{1}F_{X}^{-1}(1-u)\mathrm{d}g(u)=\int_{0}^{1}F_{X}^{-1}(u)\mathrm{d}\hat{g}(u),
\end{align*}
where $\hat{g}(u)=g(1)-g(1-u),~ u \in [0, 1]$.
\end{lemma}

\begin{remark}
Note that the similar results can be found in Lemma 3 of Wang
et al. (2020b). Their results are quantile
representation of signed Choquet integrals on $L^{\infty}$.
\end{remark}
\begin{lemma}\label{le.2}
For $g\in\mathcal{G}$, $X, \Psi(X)\in L^{0}$, weighted function $\psi(x)$: $x \in \mathbb{R}\rightarrow \psi(x)\in\mathbb{R}$ and $\psi(t)=\Psi'(t)$. \\
(i) if $g$ is right-continuous,
\begin{align}\label{a4}
   \int_{-\infty}^{+\infty}\psi(x)g(\overline{F}_{X}(x))\mathrm{d}x&=\int_{0}^{1}\Psi\left(F_{X}^{-1+}(u)\right)\mathrm{d}\hat{g}(u);
\end{align}
(ii) if $g$ is left-continuous,
\begin{align}\label{a5}
   \int_{-\infty}^{+\infty}\psi(x)g(\overline{F}_{X}(x))\mathrm{d}x&=\int_{0}^{1}\Psi\left(F_{X}^{-1}(u)\right)\mathrm{d}\hat{g}(u);
\end{align}
(iii) if $g$ is continuous,
\begin{align}\label{a6}
   \int_{-\infty}^{+\infty}\psi(x)g(\overline{F}_{X}(x))\mathrm{d}x&=\int_{0}^{1}\Psi\left(F_{X}^{-1}(u)\right)\mathrm{d}\hat{g}(u),
\end{align}
where $\hat{g}(u)=-g(1-u),~ u \in [0, 1]$.
\end{lemma}
\noindent $\mathbf{Proof.}$ ($i$) Because $g$ is right-continuous and is of bounded variation, using relations
 $g(P(X\geq x))=-\int_{(P(X\geq x),1]}\mathrm{d}g(u)$, $g(P(X\geq x))=\int_{[0,P(X\geq x)]}\mathrm{d}g(u)$ and Fubini's theorem, we get
\begin{align*}
\int_{- \infty}^{+ \infty} \psi(x)g(\overline{F}_{X}(x))\mathrm{d}x&=  \int_{- \infty}^{0}\psi(x)g(P(X\geq x)) \mathrm{d}x +
\int_{0}^{+\infty}\psi(x) g(P(X\geq x)) \mathrm{d}x\\
&= -\int_{- \infty}^{0}\psi(x)\left(\int _{(P(X\geq x),1]}\mathrm{d}g(u)\right) \mathrm{d}x+
\int_{0}^{+\infty}\psi(x) \left(\int _{[0,P(X\geq x)]}\mathrm{d}g(u)\right) \mathrm{d}x\\
&=-\int_{(P(X\geq x),1]}\left(\int _{F_{X}^{-1+}(1-u)}^{0}\psi(x)\mathrm{d}x\right) \mathrm{d}g(u) +
\int_{[F_{X}(0-),1]}\left(\int _{0}^{F_{X}^{-1+}(1-u)}\psi(x)\mathrm{d}x\right) \mathrm{d}g(u)\\
&=\int_{0}^{1}\Psi\left(F_{X}^{-1+}(1-u)\right)\mathrm{d}g(u),
\end{align*}
where the last second equality we have used  $u\leq P(X\geq x)\Leftrightarrow F_{X}^{-1+}(1-u)\geq x$.
Next, let $t=1-u$, we can get Eq. (\ref{a4})\\
$(ii)$ Similar to $(i)$'s manner, using $u\geq P(X\geq x)\Leftrightarrow F_{X}^{-1+}(1-u)\leq x$ we easily obtain Eq. (\ref{a5}).\\
$(iii)$ Similar to $(i)$ and $(ii)$'s manner, we instantly obtain Eq. (\ref{a6}),
ending the proof. $\hfill\square$
\begin{remark}
Note that the similar results can be found in Theorem 2 of Zuo and Yin (2023). However, the results of the two are mainly different in two aspects. First, the space of distortion function $g$ is different, with $g$ being of~ bounded variation and the other being continuous and almost everywhere differentiable; Second, integrand is different, with one integrand being a function of $\bar{F}$ (i.e., $g(\bar{F})$)  and the other being a function of $F$ (i.e., $g(F)$).
\end{remark}
To present the worst-cases of for distortion riskmetrics and weighted entropy, similar to Boyd and Vandenberghe (2004), we define convex and concave envelopes for distortion function $g$ as, respectively,
\begin{align*}
g_{\ast}&=\sup\left\{h|h:[0,1]\rightarrow[0,1]~ \mathrm{is} ~\mathrm{convex}~ \mathrm{and} ~h(u)\leq g(u),~u\in[0,1]\right\},\\
g^{\ast}&=\inf\left\{h|h:[0,1]\rightarrow[0,1]~ \mathrm{is}~ \mathrm{concave}~ \mathrm{and}~ h(u)\geq g(u),~u\in[0,1]\right\}.
\end{align*}
It is easy to verify $(-g_{\ast})=-g^{\ast}$. In addition, when $g$ is convex, $g_{\ast}$ is equal to $g$; when $g$ is concave, $g^{\ast}$ is equal to $g$.
\section{Main results\label{sec:3}}
In this section, we mainly derive the worst-cases of distortion riskmetrics and weighted entropy.
\begin{theorem}\label{th.1}
Under conditions of Lemma \ref{le.1}, if there are mean $\mu$ and variance $\sigma^{2}$ of random variable $X$, then we have
 \begin{align}\label{a7}
   \sup_{X\in V (\mu,\sigma)}\rho_{g}(X)= \mu g(1)+\sigma\sqrt{\int_{0}^{1}(\hat{g}_{\ast}'(u)-g(1))^{2}\mathrm{d}u},
\end{align}
where $\hat{g}(u)=g(1)-g(1-u),~ u \in [0, 1]$.
If $\hat{g}_{\ast}'(u)=g(1)$ is almost everywhere (a.e.), then (\ref{a7}) can be obtained by any random variable $X \in
V (\mu, \sigma)$; If $\hat{g}_{\ast}'(u)\neq g(1) ~(a.e.)$, then (\ref{a7})
 can be obtained by the
worst-case distribution of r.v. $X_{\ast}$ characterized by
$$F_{X_{\ast}}^{-1+}(u)=\mu+\sigma\frac{\hat{g}_{\ast}'(u)-g(1)}{\sqrt{\int_{0}^{1}\left(\hat{g}_{\ast}'(u)-g(1)\right)^{2}\mathrm{d}u}}.$$
\end{theorem}
\noindent $\mathbf{Proof.}$ When $\hat{g}_{\ast}'(u)-g(1) =0 ~(a.e.)$, the result is obvious. When $\hat{g}_{\ast}'(u)-g(1) \neq0 ~(a.e.)$, by Lemma \ref{le.1}, and using modified Schwartz inequality (see Moriguti (1953)) and Cauchy-Schwartz inequality, for any constant $c$, we get
\begin{align*}
   \rho_{g}(X)&=\mu g(1)+\int_{0}^{1}\left(F_{X}^{-1+}(u)-\mu\right)\mathrm{d}\hat{g}(u)\\
   \nonumber&\leq\mu g(1)+\int_{0}^{1}\left(F_{X}^{-1+}(u)-\mu\right)\hat{g}_{\ast}'(u)\mathrm{d}u
   \end{align*}
   \begin{align*}
   \nonumber&=\mu g(1)+\int_{0}^{1}\left(F_{X}^{-1+}(u)-\mu\right)(\hat{g}_{\ast}'(u)-c)\mathrm{d}u\\
  &\leq\mu g(1)+\left[\int_{0}^{1}\left(F_{X}^{-1+}(u)-\mu\right)^{2}\mathrm{d}u\right]^{\frac{1}{2}}\left[\int_{0}^{1}(\hat{g}_{\ast}'(u)-c)^{2}\mathrm{d}u\right]^{\frac{1}{2}}\\
   \nonumber&=\mu g(1)+\sigma\left[\int_{0}^{1}(\hat{g}_{\ast}'(u)-c)^{2}\mathrm{d}u\right]^{\frac{1}{2}}.\\
   \end{align*}
 While $$\inf_{c\in\mathbb{R}}\left[\int_{0}^{1}(\hat{g}_{\ast}'(u)-c)^{2}\mathrm{d}u\right]^{\frac{1}{2}}=\sqrt{\int_{0}^{1}(\hat{g}_{\ast}'(u)-g(1))^{2}\mathrm{d}u},$$  thus,
 \begin{align*}
 \rho_{g}(X)\leq \mu g(1)+\sigma\sqrt{\int_{0}^{1}(\hat{g}_{\ast}'(u)-g(1))^{2}\mathrm{d}u},
 \end{align*}
 where equality holds if and only if
$$F_{X_{\ast}}^{-1+}(u)=\mu+\sigma\frac{\hat{g}_{\ast}'(u)-g(1)}{\sqrt{\int_{0}^{1}\left(\hat{g}_{\ast}'(u)-g(1)\right)^{2}\mathrm{d}u}}.$$
In fact, it is easy to verify that
 \begin{align*}
\int_{0}^{1}\left(F_{X_{\ast}}^{-1+}(u)-\mu\right)\mathrm{d}\hat{g}(u)=\frac{\sigma\int_{0}^{1}(\hat{g}_{\ast}'(u)-g(1))\mathrm{d}(\hat{g}(u)-g(1)u)}{\sqrt{\int_{0}^{1}(\hat{g}_{\ast}'(u)-g(1))^{2}\mathrm{d}u}}=\sigma\sqrt{\int_{0}^{1}(\hat{g}_{\ast}'(u)-g(1))^{2}\mathrm{d}u},
 \end{align*}
 where the second equality we have used the modified Schwartz inequality again.
Therefore, (\ref{a7}) holds, which completes the proof.  $\hfill\square$
\begin{remark}
Note that when $g$ is a nondecreasing function on $[0, 1]$ and $g(1) = 1$, Theorem \ref{th.1} will be reduced to Proposition 4.1 of  Zhao et al. (2024).
\end{remark}
\begin{corollary}\label{co.1}
 Let $g(1) = 0$ in Theorem \ref{th.1}, worst-case of entropy is given by
\begin{align}\label{a8}
   \sup_{X\in V (\mu,\sigma)}\int_{-\infty}^{+\infty}g(\overline{F}_{X}(x))\mathrm{d}x= \sigma\sqrt{\int_{0}^{1}(\hat{g}_{\ast}'(u))^{2}\mathrm{d}u},
\end{align}
where $\hat{g}(u)=-g(1-u)$, $u\in[0,1]$.
If $\hat{g}_{\ast}'(u)=0~(a.e.)$, then (\ref{a8}) can be obtained by any random variable $X \in
V (\mu, \sigma)$; If $\hat{g}_{\ast}'(u)\neq 0 ~(a.e.)$, then (\ref{a8})
 can be obtained by the
worst-case distribution of r.v. $X_{\ast}$ characterized by
$$F_{X_{\ast}}^{-1+}(u)=\mu+\sigma\frac{\hat{g}_{\ast}'(u)}{\sqrt{\int_{0}^{1}\left(\hat{g}_{\ast}'(u)\right)^{2}\mathrm{d}u}}.$$
\end{corollary}

\begin{remark}
 Note that when $g(u)=-h'(u)(u)\log(u)$ in Corollary \ref{co.1}, the above entropy (form) will be reduced to $h$-CRE
family of variability measures defined by Psarrakos et al. (2024).
\end{remark}

\begin{theorem}\label{th.2}
Under conditions of Lemma \ref{le.2}, if there are mean $\mu_{\Psi}=\mathrm{E}[\Psi(X)]$ and variance $\sigma_{\Psi}^{2}=\mathrm{E}[\Psi(X)-\mu_{\Psi}]^{2}$ of random variable $\Psi(X)$, then we have
 \begin{align}\label{a9}
   \sup_{X\in V (\mu_{\Psi},\sigma_{\Psi})}\int_{-\infty}^{+\infty}\psi(x)g(\overline{F}_{X}(x))\mathrm{d}x= \sigma_{\Psi}\sqrt{\int_{0}^{1}\left(\hat{g}_{\ast}'(u)\right)^{2}\mathrm{d}u},
\end{align}
where $\hat{g}(u)=-g(1-u)$, $u\in[0,1]$.
If $\hat{g}_{\ast}'(u) =0 ~(a.e.)$, then (\ref{a9}) can be obtained by any random variable $X \in
V (\mu_{\Psi}, \sigma_{\Psi})$; If $\hat{g}_{\ast}'(u) \neq0 ~(a.e.)$, then (\ref{a9})
 can be obtained by the
worst-case distribution of r.v. $X_{\ast}$ characterized by
$$\Psi(F_{X_{\ast}}^{-1+}(u))=\mu_{\Psi}+\sigma_{\Psi}\frac{\hat{g}_{\ast}'(u)}{\sqrt{\int_{0}^{1}\left(\hat{g}_{\ast}'(u)\right)^{2}\mathrm{d}u}}.$$
\end{theorem}
\noindent $\mathbf{Proof.}$ When $\hat{g}_{\ast}'(u)=0 ~(a.e.)$, the result is obvious. When $\hat{g}_{\ast}'(u) \neq0 ~(a.e.)$, by Lemma \ref{le.2}, and using the modified Schwartz inequality (see Moriguti (1953)) and the Cauchy-Schwartz inequality, for any constant $c$, we get
\begin{align*}
   \nonumber\int_{-\infty}^{+\infty}\psi(x)g(\overline{F}_{X}(x))\mathrm{d}x&=\int_{0}^{1}\Psi\left(F_{X}^{-1+}(u)\right)\mathrm{d}\hat{g}(u)\\
   &=\int_{0}^{1}\left(\Psi\left(F_{X}^{-1+}(u)\right)-\mu_{\Psi}\right)\mathrm{d}\hat{g}(u)\\
   \nonumber&\leq\int_{0}^{1}\left(\Psi\left(F_{X}^{-1+}(u)\right)-\mu_{\Psi}\right)\hat{g}_{\ast}'(u)\mathrm{d}u\\
   \nonumber&=\int_{0}^{1}\left(\Psi\left(F_{X}^{-1+}(u)\right)-\mu_{\Psi}\right)(\hat{g}_{\ast}'(u)-c)\mathrm{d}u\\
  &\leq\left[\int_{0}^{1}\left(\Psi\left(F_{X}^{-1+}(u)\right)-\mu_{\Psi}\right)^{2}\mathrm{d}u\right]^{\frac{1}{2}}\left[\int_{0}^{1}(\hat{g}_{\ast}'(u)-c)^{2}\mathrm{d}u\right]^{\frac{1}{2}}\\
   \nonumber&=\sigma_{\Psi}\left[\int_{0}^{1}\left(\hat{g}_{\ast}'(u)-c\right)^{2}\mathrm{d}u\right]^{\frac{1}{2}}\\
   \end{align*}
While
$$\inf_{c\in\mathbb{R}}\left[\int_{0}^{1}(\hat{g}_{\ast}'(u)-c)^{2}\mathrm{d}u\right]^{\frac{1}{2}}=\sqrt{\int_{0}^{1}\left(\hat{g}_{\ast}'(u)\right)^{2}\mathrm{d}u},$$
thus we have
\begin{align*}
   \int_{-\infty}^{+\infty}\psi(x)g(\overline{F}_{X}(x))\mathrm{d}x\leq \sigma_{\Psi}\sqrt{\int_{0}^{1}\left(\hat{g}_{\ast}'(u)\right)^{2}\mathrm{d}u},
\end{align*}
where equality holds if and only if
$$\Psi(F_{X_{\ast}}^{-1+}(u))=\mu_{\Psi}+\sigma_{\Psi}\frac{\hat{g}_{\ast}'(u)}{\sqrt{\int_{0}^{1}\left(\hat{g}_{\ast}'(u)\right)^{2}\mathrm{d}u}}.$$
In this case, it is easy to verify that
\begin{align*}
\int_{0}^{1}\left(\Psi(F_{X_{\ast}}^{-1+}(u))-\mu_{\Psi}\right)\mathrm{d}\hat{g}(u)=\sigma_{\Psi}\sqrt{\int_{0}^{1}\left(\hat{g}_{\ast}'(u)\right)^{2}\mathrm{d}u},
 \end{align*}
 where we have used the modified Schwartz inequality again in the equality here.
Therefore, (\ref{a9}) holds, completing the proof.  $\hfill\square$
\begin{remark}
Note that let $\psi(x)=1$ in Theorem \ref{th.2}, we can obtain the worst-case of entropy as the result in Corollary \ref{co.1}.
\end{remark}
For a random variable $X_{t}=[X-t|X>t]$, it's survival function is
 \begin{align*}
 \overline{F}_{X_{t}}(x)=
 \begin{cases}
         \frac{\overline{F}_{X}(x)}{\overline{F}_{X}(t)},~\mathrm{when} ~x>t,\\
      1              ,~\mathrm{otherwise}.
 \end{cases}
 \end{align*}
 Thus, for any $v \in (0, 1)$, $F_{X_{t}}^{-1}(x)=F_{X}^{-1}\left(F_{X}(t)+(1-F_{X}(t))v\right).$

\begin{corollary}\label{co.2}
Let $X=X_{t}$ in
Theorem \ref{th.2}, then we get
 \begin{align}\label{a10}
   \sup_{X\in V (\mu_{\Psi},\sigma_{\Psi})}\int_{-\infty}^{+\infty}\psi(x)g(\overline{F}_{X_{t}}(x))\mathrm{d}x= \sigma_{\Psi}\sqrt{\int_{0}^{1}\left([\hat{g}_{t}(u)]_{\ast}'\right)^{2}\mathrm{d}u},
\end{align}
where $\hat{g}_{t}(u)=-g\left(\frac{1-u}{1-F_{X}(t)}\right)\mathbb{I}_{[F_{X}(t),1]}(u))$, $u\in[0,1]$.
If $[\hat{g}_{t}(u)]_{\ast}' =0 ~(a.e.)$, then (\ref{a10}) can be obtained by any random variable $X \in
V (\mu_{\Psi},\sigma_{\Psi})$; If $[\hat{g}_{t}(u)]_{\ast}' \neq0 ~(a.e.)$, then (\ref{a10})
 can be obtained by the
worst-case distribution of r.v. $X_{\ast}$ characterized by
$$\Psi(F_{X_{\ast}}^{-1+}(u))=\mu_{\Psi}+\sigma_{\Psi}\frac{(-g_{t}(1-u))_{\ast}'}{\sqrt{\int_{0}^{1}\left((-g_{t}(1-u))_{\ast}'\right)^{2}\mathrm{d}u}}.$$
\end{corollary}
For a random variable $X_{(t)}=[X|X\leq t]$, it's distribution function is represented as
 \begin{align*}
 F_{X_{(t)}}(x)=
 \begin{cases}
         \frac{F_{X}(x)}{F_{X}(t)},~\mathrm{when} ~x\leq t,\\
      0              ,~\mathrm{otherwise}.
 \end{cases}
 \end{align*}
Thus, for any $s \in (0, 1)$, $F_{X_{(t)}}^{-1}(x)=F_{X}^{-1}\left(F_{X}(t)s\right).$
\begin{corollary}\label{co.3}
Let $X=X_{(t)}$ in
Theorem \ref{th.2}, then we obtain
 \begin{align}\label{a11}
   \sup_{X\in V (\mu_{\Psi},\sigma_{\Psi})}\int_{-\infty}^{+\infty}\psi(x)g(\overline{F}_{X_{(t)}}(x))\mathrm{d}x= \sigma_{\Psi}\sqrt{\int_{0}^{1}\left([\hat{g}_{(t)}(u)]_{\ast}'\right)^{2}\mathrm{d}u},
\end{align}
where
 $\hat{g}_{(t)}(u)=-g\left(\frac{u}{F_{X}(t)}\right)\mathbb{I}_{[0,F_{X}(t)]}(u)$, $u\in[0,1]$.
If $[\hat{g}_{(t)}(u)]_{\ast}' =0 ~(a.e.)$, then (\ref{a11}) can be obtained by any random variable $X \in
V (\mu_{\Psi},\sigma_{\Psi})$; If $[\hat{g}_{(t)}(u)]_{\ast}' \neq0 ~(a.e.)$, then (\ref{a11})
 can be obtained by the
worst-case distribution of r.v. $X_{\ast}$ characterized by
$$\Psi(F_{X_{\ast}}^{-1+}(u))=\mu_{\Psi}+\sigma_{\Psi}\frac{[\hat{g}_{(t)}(u)]_{\ast}'}{\sqrt{\int_{0}^{1}\left((-g_{(t)}(1-u))_{\ast}'\right)^{2}\mathrm{d}u}}.$$
\end{corollary}
\section{Applications to Entropies\label{sec:4}}
In this section, we apply the results of the previous section to some the commonly used entropies, such as Gini, (dynamic) CRE, TGini, (dynamic) CT and (tail) EGini.

%%%%%%%%%%%%%%%%%%%%%%%%%%
\begin{example}\label{ex.1} The cumulative Tsallis past entropy of a random variable $X$, with distribution function $F_{X}(x)$, denoted by $\mathrm{CT}_{\alpha}(X)$ (when $X\in L^{0}_{+}$, see Cal\`{\i} et al. (2017)), is given by
\begin{align*} \mathrm{CT}_{\alpha}(X)&=\int_{-\infty}^{+\infty}\frac{1}{\alpha-1}[F_{X}(x)-(F_{X}(x))^{\alpha}]\mathrm{d}x.
\end{align*}
In this case, $g(u)=\frac{1}{\alpha-1}[(1-u)-(1-u)^{\alpha}], ~\alpha>0,~\alpha\neq1,~u\in[0,1]$ in Corollary \ref{co.1}. Because $\hat{g}(u)=-g(1-u)$ is convex in $[0,1]$, $[\hat{g}(u)]_{\ast}=-g(1-u)$ (for instance, $\alpha=\frac{2}{3},2 ~\mathrm{and} ~3$, see Fig. 1(a)). Then,
\begin{align*}
\sup_{X\in V (\mu,\sigma)} \mathrm{CT}_{\alpha}(X)&=\frac{\sigma}{\sqrt{2\alpha-1}},~\alpha>\frac{1}{2},
\end{align*}
which can be obtained by the
worst-case distribution of r.v. $X_{\ast}$ characterized by
$$F_{X_{\ast}}^{-1}(u)=\mu+\sigma\frac{\sqrt{2\alpha-1}}{\alpha-1}\left(\alpha u^{\alpha-1}-1\right).$$
\end{example}

When $\alpha = 2$, $\mathrm{CT}_{\alpha}(X)$ is reduced to Gini mean semi-difference ($\mathcal{G}ini(X)$) (see Hu and Chen (2020)), defined by
\begin{align*}
\mathcal{G}ini(X)&=\int_{-\infty}^{+\infty}[F_{X}(x)-(F_{X}(x))^{2}]\mathrm{d}x.
\end{align*}
In this case, $g(u) = u(1 - u)$, $u\in[0,1]$. Then, we get
$$\sup_{X\in V (\mu,\sigma)} \mathcal{G}ini(X)=\frac{\sigma}{\sqrt{3}},$$
which can be obtained by the
worst-case distribution of r.v. $X_{\ast}$ characterized by
 $$F_{X_{\ast}}^{-1}(u)=\mu+\sqrt{3}(2u-1)\sigma.$$

 \begin{example}\label{ex.2}A random variable $X$, with distribution function $F_{X}(x)$ and tail distribution function $\overline{F}_{X}(x)$, has its
 the  cumulative residual Tsallis entropy of order $\alpha$, denoted by $\mathrm{CRT}_{\alpha}(X)$ (when $X\in L^{0}_{+}$, see Rajesh and Sunoj (2019)), as
\begin{align*}
\mathrm{CRT}_{\alpha}(X)
&=\int_{-\infty}^{+\infty}\frac{1}{\alpha-1}\left[\overline{F}_{X}(x)-(\overline{F}_{X}(x))^{\alpha}\right]\mathrm{d}x.
\end{align*}
In this case,  $g(u)=\frac{1}{\alpha-1}[u-u^{\alpha}], ~\alpha>0,~\alpha\neq1,~u\in[0,1]$ in Corollary \ref{co.1}. Because $\hat{g}(u)=-g(1-u)$ is convex in $[0,1]$, $\hat{g}_{\ast}(u)=-g(1-u)$ (for instance, $\alpha=\frac{2}{3},\rightarrow1,=2$, see Fig. 1(b)). Then,
\begin{align*}
\sup_{X\in V (\mu,\sigma)} \mathrm{CRT}_{\alpha}(X)&=\frac{\sigma}{\sqrt{2\alpha-1}},~\alpha>\frac{1}{2},
\end{align*}
which can be obtained by the
worst-case distribution of r.v. $X_{\ast}$ characterized by
$$F_{X_{\ast}}^{-1}(u)=\mu+\sigma\frac{\sqrt{2\alpha-1}}{\alpha-1}\left[1-\alpha(1-u)^{\alpha-1}\right].$$
 \end{example}

\begin{example}\label{ex.3} The extended Gini coefficient of a random variable $X$, with distribution function $F_{X}(x)$ and tail distribution function $\overline{F}_{X}(x)$, denoted by $\mathrm{EGini}_{r}(X)$ (when $X\in L^{0}_{+}$, see Berkhouch et al. (2018)), is given by
\begin{align*}
 \mathrm{EGini}_{r}(X)&=-2\int_{-\infty}^{+\infty}\left[F_{X}(x)+\left(\overline{F}_{X}(x)\right)^{r}-1\right]\mathrm{d}x.
 \end{align*}
In this case, $g(u)=-2[1-u+u^{r}-1],~r>1,~u\in[0,1]$ in Corollary \ref{co.1}. Because $\hat{g}(u)=-g(1-u)$ is convex in $[0,1]$, $\hat{g}_{\ast}(u)=-g(1-u)$. For instance,  $r=\frac{3}{2},2,3,$ $-g(1-u)$ is plotted in Fig. 1 (c). Then, we attain
\begin{align*}
\sup_{X\in V (\mu,\sigma)}\mathrm{EGini}_{r}(X)&=\frac{2(r-1)}{\sqrt{2r-1}}\sigma,
\end{align*}
which can be obtained by the
worst-case distribution of r.v. $X_{\ast}$ characterized by
$$F_{X_{\ast}}^{-1}(u)=\mu+\sigma\frac{\sqrt{2r-1}}{r-1}\left[1-r(1-u)^{r-1}\right].$$
\end{example}

Note that when $r=2$, the extended Gini coefficient reduces to the simple Gini ($\mathrm{Gini}(X)$) (see Furman et al. (2017)). Then,
\begin{align*}
\sup_{X\in V (\mu,\sigma)}\mathrm{Gini}(X)= \sup_{X\in V (\mu,\sigma)}\mathrm{EGini}_{2}(X)&=\frac{2}{\sqrt{3}}\sigma,
\end{align*}
which can be obtained by the
worst-case distribution of r.v. $X_{\ast}$ characterized by
$$F_{X_{\ast}}^{-1}(u)=\mu+\sqrt{3}\left[1-2(1-u)\right]\sigma.$$

%%%%%%%%%%%%%%%%%%%%%%%%%%

\begin{example}\label{ex.4} A random variable $X$, with distribution function $F_{X}(x)$ and tail distribution function $\overline{F}_{X}(x)$, has its
 the  fractional generalized cumulative residual entropy, denoted by $FGR\mathcal{E}_{\alpha}(X)$ (when $X\in(0,c)$, see Di Crescenzo et al. (2021)), as
\begin{align*}
 FGR\mathcal{E}_{\alpha}(X)& =\frac{1}{\Gamma(\alpha+1)}\int_{-\infty}^{+\infty}\overline{F}_{X}(x)\left[-\log\left(\overline{F}_{X}(x)\right)\right]^{\alpha}\mathrm{d}x.
 \end{align*}
In this case, $g(u)=\frac{1}{\Gamma(\alpha+1)}u[-\log(u)]^{\alpha}, ~\alpha>0,~u\in[0,1]$ in Corollary \ref{co.1}.
(i) When $\alpha\in(0,1],$ $\hat{g}(u)=-g(1-u)$ is convex in $[0,1]$, $\hat{g}_{\ast}(u)=-g(1-u)$, for instance, $\alpha=\frac{1}{2},\frac{2}{3} ~\mathrm{and}~ 1$, plotted in Fig. 2(a). So that (For $X \in (0, c)$ and $\alpha\in(0,1]$, see Xiong et al. (2019))
 \begin{align*}
\sup_{X\in V (\mu,\sigma)}FGR\mathcal{E}_{\alpha}(X)&=\sigma\frac{\sqrt{\Gamma(2\alpha-1)}}{\Gamma(\alpha)},\alpha>\frac{1}{2},
\end{align*}
which can be obtained by the
worst-case distribution of r.v. $X_{\ast}$ characterized by
$$F_{X_{\ast}}^{-1}(u)=\mu+\sigma\frac{[-g(1-u)]'\Gamma(\alpha)}{\sqrt{\Gamma(2\alpha-1)}}.$$

%%%%%%%%%%%%%%%%%%%%%%%%%%%%%%%%%%%%%%%%%
%%%%%%%%%%%%%%%%%%%%%%%%%%%%%%%%%%%%%%%%5
(ii) When $\alpha\in(1,\infty),$
because $\hat{g}(u)=-g(1-u)$ is concave in $[0,1-e^{1-\alpha})$, and is convex in $[1-e^{1-\alpha},1]$, for instance, $\alpha=1,2 ~\mathrm{and}~ 3$, plotted in Fig. 2(b). Then, we get\\
\begin{align*}
 \hat{g}_{\ast}(u)=[-g(1-u)]_{\ast}=
 \begin{cases}
 b u,~u\in[0,u_{0}),\\
 \frac{-1}{\Gamma(\alpha+1)}(1-u)[-\log(1-u)]^{\alpha},~u\in[u_{0},1],
 \end{cases}
 \end{align*}
 where $b=\frac{-1}{u_{0}\Gamma(\alpha+1)}(1-u_{0})[-\log(1-u_{0})]^{\alpha}$, and $u_{0}\in[1-e^{1-\alpha},1]$ is the solution to the following equation: $\alpha u_{0}+\log(1-u_{0})=0$. For example, $\alpha= 3$, $u_{0}\approx0.94048$, $-g(1-u)~ \mathrm{and} ~[-g(1-u)]_{\ast}$ are plotted in Fig. 2(c).
 Thus,
\begin{align*}
\sup_{X\in V (\mu,\sigma)}FGR\mathcal{E}_{\alpha}(X)
&=\sigma\frac{\sqrt{d}}{\Gamma(\alpha+1)},
\end{align*}
which can be obtained by the
worst-case distribution of r.v. $X_{\ast}$ characterized by
 \begin{align*}
F_{X_{\ast}}^{-1}(u)=
\begin{cases}
\mu+\sigma\frac{b\Gamma(\alpha+1)}{\sqrt{d}},~u\in[0,u_{0}),\\
\mu+\sigma\frac{[-g(1-u)]'\Gamma(\alpha+1)}{\sqrt{d}},~u\in[u_{0},1],
\end{cases}
\end{align*}
where $d=(\log(1-u_{0}))^{2\alpha}\frac{1-u_{0}}{u_{0}}+\alpha^{2}\Gamma(2\alpha-1,-\log(1-u_{0}))$,
 and $\Gamma(\cdot,\cdot)$ denotes upper incomplete gamma function (for instance, see Abramowitz and Stegun (1965), Chapter 6), i.e.,
$$\Gamma(s,x)=\int_{x}^{+\infty}t^{s-1}\mathrm{e}^{-t}\mathrm{d}t.$$
\end{example}
Note that if $\alpha = n \in N$, $FGR\mathcal{E}_{\alpha}(X)$ will become generalized cumulative residual entropy ($\mathrm{GCRE}_{n}(X)$) introduced by Psarrakos and Navarro (2013), i.e.,
\begin{align*}
 \mathrm{GCRE}_{n}(X)
 &=\frac{1}{n!}\int_{0}^{+\infty}\overline{F}_{X}(x)\left[-\log\left(\overline{F}_{X}(x)\right)\right]^{n}\mathrm{d}x.
\end{align*}
 Then, we have
\begin{align*}
\sup_{X\in V (\mu,\sigma)}\mathrm{GCRE}_{n}(X)&=\sigma\frac{\sqrt{d}}{n!},
\end{align*}
which can be obtained by the
worst-case distribution of r.v. $X_{\ast}$ characterized by
\begin{align*}
F_{X_{\ast}}^{-1}(u)=
\begin{cases}
\mu+\sigma\frac{b (n!)}{\sqrt{d}},~u\in[0,u_{0}),\\
\mu+\sigma\frac{[-g(1-u)]' (n!)}{\sqrt{d}},~u\in[u_{0},1],
\end{cases}
\end{align*}
$b=\frac{-1}{u_{0}n!}(1-u_{0})[-\log(1-u_{0})]^{n}$,
$d=(\log(1-u_{0}))^{2n}\frac{1-u_{0}}{u_{0}}+n^{2}\Gamma(2n-1,-\log(1-u_{0}))$, and $u_{0}$ is the solution to the following equation: $n u_{0}+\log(1-u_{0})=0$.

%%%%%%%%%%%%%%%%%%%%%%%%%%%%
%%%%%%%%%%%%%%%%%%%%%%%%%%%%

Note that when $\alpha= 1$, $FGR\mathcal{E}_{\alpha}(X)$ is reduced to cumulative residual entropy ($\mathcal{E}(X)$), defined as (see Rao et al. (2004))
\begin{align*}
\mathcal{E}(X)
&=-\int_{0}^{+\infty}\overline{F}_{X}(x)\mathrm{log}(\overline{F}_{X}(x))\mathrm{d}x.
\end{align*}
In this case, $g(u)=-u\mathrm{log}(u)$, $u\in[0,1]$ in Corollary \ref{co.1}. Since $\hat{g}(u)=-g(1-u)$ is convex in $[0,1]$, $\hat{g}_{\ast}(u)=-g(1-u)$ (for detail, see Fig. 2(a) $\alpha=1$). Then, we have
\begin{align*}
\sup_{X\in V (\mu,\sigma)} \mathcal{E}(X)&=\sigma,
\end{align*}
which can be obtained by the
worst-case distribution of r.v. $X_{\ast}$ characterized by
$$F_{X_{\ast}}^{-1}(u)=\mu-\sigma(\log (1-u)+1).$$
\begin{example}\label{ex.5} The fractional generalized cumulative entropy of a random variable $X$, with distribution function $F_{X}(x)$, denoted by $FG\mathcal{E}_{\alpha}(X)$ (when  $X\in(0,c)$, see Di Crescenzo et al. (2021)), given by
\begin{align*}
 FG\mathcal{E}_{\alpha}(X)
 &=\frac{1}{\Gamma(\alpha+1)}\int_{-\infty}^{+\infty}F_{X}(x)\left[-\log\left(F_{X}(x)\right)\right]^{\alpha}\mathrm{d}x.
\end{align*}
In this case, $g(u)=\frac{1}{\Gamma(\alpha+1)}(1-u)[-\log (1-u)]^{\alpha}, ~\alpha>0,~u\in[0,1]$ in Corollary \ref{co.1}.
(i) When $\alpha\in(0,1],$ $\hat{g}(u)=-g(1-u)$ is convex in $u\in[0,1]$, for instance, $\alpha=1,2 ~\mathrm{and}~ 3$, plotted in Fig. 3(a), so that $\hat{g}_{\ast}(u)=-g(1-u),$
 then,
 \begin{align*}
\sup_{X\in V (\mu,\sigma)}FG\mathcal{E}_{\alpha}(X)&=\sigma\frac{\sqrt{\Gamma(2\alpha-1)}}{\Gamma(\alpha)},~\alpha>\frac{1}{2},
\end{align*}
which can be obtained by the
worst-case distribution of r.v. $X_{\ast}$ characterized by
$$F_{X_{\ast}}^{-1}(u)=\mu+\sigma\frac{[-g(1-u)]'\Gamma(\alpha)}{\sqrt{\Gamma(2\alpha-1)}}.$$

%%%%%%%%%%%%%%%%%%%%%%%%%%%%%%%%%%%
%%%%%%%%%%%%%%%%%%%%%%%%%%%%%%%%%%%
(ii) When $\alpha\in(1,\infty),$ because $\hat{g}(u)=-g(1-u)$ is convex in $[0,e^{1-\alpha}]$, and is concave in $(e^{1-\alpha},1]$, for instance, $\alpha=1,2 ~\mathrm{and}~ 3$, plotted in Fig. 3(b), then,
\begin{align*}
 \hat{g}_{\ast}(u)=[-g(1-u)]_{\ast}=
 \begin{cases}
 \frac{-1}{\Gamma(\alpha+1)}u[-\log u]^{\alpha},~u\in[0,u_{1}],\\
 b_{1} u-b_{1},~u\in(u_{1},1],
 \end{cases}
 \end{align*}
 where $b_{1}=\frac{u_{1}[-\log(u_{1})]^{\alpha}}{(1-u_{1})\Gamma(\alpha+1)}$, and $u_{1}\in[0,e^{1-\alpha}]$ is the solution to the following equation: $\alpha (1-u_{1})+\log(u_{1})=0$. For example, $\alpha= 3$, $u_{1}\approx0.05952$, $-g(1-u)~ \mathrm{and} ~[-g(1-u)]_{\ast}$ are plotted in Fig. 3(c).
 Hence,
\begin{align*}
\sup_{X\in V (\mu,\sigma)}FG\mathcal{E}_{\alpha}(X)&=\sigma\frac{\sqrt{d}}{\Gamma(\alpha+1)},
\end{align*}
which can be obtained by the
worst-case distribution of r.v. $X_{\ast}$ characterized by
 \begin{align*}
F_{X_{\ast}}^{-1}(u)=
\begin{cases}
\mu+\sigma\frac{[-g(1-u)]'\Gamma(\alpha+1)}{\sqrt{d}},~u\in[0,u_{1}],\\
\mu+\sigma\frac{b_{1} \Gamma(\alpha+1)}{\sqrt{d}},~u\in(u_{1},1],
\end{cases}
\end{align*}
with $d=(\log(u_{1}))^{2\alpha}\frac{u_{1}}{1-u_{1}}+\alpha^{2}\Gamma(2\alpha-1,-\log(u_{1}))$.
\end{example}
Note that if $\alpha = n \in N$, $FG\mathcal{E}_{\alpha}(X)$ will reduce to generalized cumulative entropy ($\mathrm{GCE}_{n}(X)$) introduced by Kayal (2016) (see also Cal\`{\i} et al. (2020)), i.e.,
\begin{align*}
 \mathrm{GCE}_{n}(X)&=\frac{1}{n!}\int_{0}^{+\infty}F_{X}(x)\left[-\log\left(F_{X}(x)\right)\right]^{n}\mathrm{d}x.
\end{align*}
Thus,
\begin{align*}
\sup_{X\in V (\mu,\sigma)}\mathrm{GCE}_{n}(X)&=\sigma\frac{\sqrt{d}}{n!},
\end{align*}
which can be obtained by the
worst-case distribution of r.v. $X_{\ast}$ characterized by
\begin{align*}
F_{X_{\ast}}^{-1}(u)=
\begin{cases}
\mu+\sigma\frac{[-g(1-u)]'(n!)}{\sqrt{d}},~u\in[0,u_{1}],\\
\mu+\sigma\frac{b_{1} (n!)}{\sqrt{d}},~u\in(u_{1},1],
\end{cases}
\end{align*}
$b_{1}=\frac{u_{1}[-\log(u_{1})]^{\alpha}}{(1-u_{1})n!}$,
$d=(\log(u_{1}))^{2n}\frac{u_{1}}{1-u_{1}}+n^{2}\Gamma(2n-1,-\log(u_{1}))$, and $u_{1}$ is the solution to the following equation: $n (1-u_{1})+\log(u_{1})=0$.

Note that when $\alpha= 1$, $FG\mathcal{E}_{\alpha}(X)$ is reduced to cumulative entropy, denoted by $\mathcal{CE}(X)$, defined as (see Di Crescenzo and Longobardi (2009))
\begin{align*}
\mathcal{CE}(X)
&=-\int_{0}^{+\infty}F_{X}(x)\mathrm{log}(F_{X}(x))\mathrm{d}x.
\end{align*}
In this case, $g(u)=-(1-u)\mathrm{log}(1-u)$ in Corollary \ref{co.1}. Since $\hat{g}(u)=-g(1-u)$ is convex in $[0,1]$, $\hat{g}_{\ast}(u)=-g(1-u)$ (for detail, see Fig. 3(a) $\alpha=1$). Then, we have
\begin{align*}
\sup_{X\in V (\mu,\sigma)} \mathcal{CE}(X)&=\sigma,
\end{align*}
which can be obtained by the
worst-case distribution of r.v. $X_{\ast}$ characterized by
$$F_{X_{\ast}}^{-1}(u)=\mu+\sigma(\log u+1).$$

%%%%%%%%%%%%%%%%%%%%%%%%%%

\begin{example}\label{ex.6} A random variable $X$, with distribution function $F_{X}(x)$ and tail distribution function $\overline{F}_{X}(x)$, has its
 the dynamic cumulative residual Tsallis entropy of order $\alpha$,
  denoted by $\mathrm{DCRT}_{\alpha,t}(X)$ (for $X\in L_{+}^{0}$, see Rajesh and Sunoj (2019)), as
 \begin{align*}
 \mathrm{DCRT}_{\alpha,t}(X)
&=\int_{t}^{+\infty}\frac{1}{\alpha-1}\left[\frac{\overline{F}_{X}(x)}{\overline{F}_{X}(t)}-\left(\frac{\overline{F}_{X}(x)}{\overline{F}_{X}(t)}\right)^{\alpha}\right]\mathrm{d}x.
\end{align*}
In this case, $\psi(x)=1$ and $g(u)=\frac{1}{\alpha-1}(u-u^{\alpha})$, $\alpha>0,~\alpha\neq1$, in Corollary \ref{co.2}. Since $\hat{g}_{t}(u)=-\frac{1}{\alpha-1}\left[\frac{1-u}{1-F_{X}(t)}-\left(\frac{1-u}{1-F_{X}(t)}\right)^{\alpha}\right]\mathbb{I}_{[F_{X}(t),1]}(u)$ is not convex in $[0,1]$,
 \begin{align*}
 [\hat{g}_{t}(u)]_{\ast}=
 \begin{cases}
 b u,~u\in[0,u_{0}],\\
 -\frac{1}{\alpha-1}\left[\frac{1-u}{1-F_{X}(t)}-\left(\frac{1-u}{1-F_{X}(t)}\right)^{\alpha}\right],~u\in(u_{0},1],
 \end{cases}
 \end{align*}
 where $b=\frac{-1}{(\alpha-1)u_{0}}\left[\frac{1-u_{0}}{1-F_{X}(t)}-\left(\frac{1-u_{0}}{1-F_{X}(t)}\right)^{\alpha}\right]$, and $u_{0}\in[F_{X}(t),1]$ is the solution to the following equation: $(1-F_{X}(t))^{\alpha-1}-(1-u_{0})^{\alpha-1}[1+(\alpha-1)u_{0}]=0$. For instance, $\alpha=2$ and $F_{X}(t)=0.2$, then $u_{0}=\sqrt{0.2}$, $\hat{g}_{t}(u)~ \mathrm{and} ~[\hat{g}_{t}(u)]_{\ast}$ are plotted in Fig. 4(a).
 So that
 \begin{align*}
 \sup_{X\in V (\mu,\sigma)} \mathrm{DCRT}_{\alpha,t}(X)&=\frac{\sigma\sqrt{d}}{|\alpha-1|(1-F_{X}(t))},~\alpha>\frac{1}{2},
 \end{align*}
which can be obtained by the
worst-case distribution of r.v. $X_{\ast}$ characterized by
\begin{align*}
F_{X_{\ast}}^{-1}(u)=
\begin{cases}
\mu+\sigma\frac{\mathrm{sign}(\alpha-1)}{\sqrt{d}}\left[1-\alpha\left(\frac{u-F_{X}(t)}{1-F_{X}(t)}\right)^{\alpha-1}\right],~u\in[u_{0},1],\\
\mu+\sigma\frac{|\alpha-1|(1-F_{X}(t))b}{\sqrt{d}},~u\in[0,u_{0}),
\end{cases}
\end{align*}
where $d=\frac{1-u_{0}}{u_{0}}-\frac{2(1-u_{0})^{\alpha}}{u_{0}(1-F_{X}(t))^{\alpha-1}}+\frac{(1-u_{0})^{2\alpha-1}}{(1-F_{X}(t))^{2\alpha-2}}\left[\frac{1}{u_{0}}+\frac{(\alpha-1)^{2}}{2\alpha-1}\right]$, and $\mathrm{sign}(\cdot)$ and $|\cdot|$ are the sign and absolute value functions, respectively.
\end{example}

When $t=x_{p}$, above $\mathrm{DCRT}_{\alpha,t}(X)$ will be tail-based cumulative residual Tsallis entropy of order $\alpha$, denoted by $\mathrm{TCRTE}_{\alpha,p}(X)$, for instance, see Zuo and Yin (2023), i.e.,
\begin{align}\label{a12}
\mathrm{TCRTE}_{\alpha,p}(X)
=\int_{x_{p}}^{+\infty}\frac{1}{\alpha-1}\left[\frac{\overline{F}_{X}(x)}{1-p}-\left(\frac{\overline{F}_{X}(x)}{1-p}\right)^{\alpha}\right]\mathrm{d}x.
\end{align}
In this case,  $\hat{g}_{t}(u)=\hat{g}_{\alpha,p}(u)=-\frac{1}{\alpha-1}\left[\frac{1-u}{1-p}-\left(\frac{1-u}{1-p}\right)^{\alpha}\right]\mathbb{I}_{[p,1]}(u)$, $\alpha>0,~\alpha\neq1$, $u\in[0,1]$, and
\begin{align*}
 [\hat{g}_{\alpha,p}(u)]_{\ast}=
 \begin{cases}
 b u,~u\in[0,u_{0}],\\
 -\frac{1}{\alpha-1}\left[\frac{1-u}{1-p}-\left(\frac{1-u}{1-p}\right)^{\alpha}\right],~u\in(u_{0},1],
 \end{cases}
 \end{align*}
 where $b=\frac{-1}{(\alpha-1)u_{0}}\left[\frac{1-u_{0}}{1-p}-\left(\frac{1-u_{0}}{1-p}\right)^{\alpha}\right]$, and $u_{0}\in[p,1]$ is the solution to the following equation: $(1-p)^{\alpha-1}-(1-u_{0})^{\alpha-1}[1+(\alpha-1)u_{0}]=0$. For instance, $\alpha=3$ and $p=0.5$, then $u_{0}\approx0.67365$, $\hat{g}_{\alpha,p}(u)~ \mathrm{and} ~[\hat{g}_{\alpha,p}(u)]_{\ast}$ are plotted in Fig. 4(b).
 Thus, we have
\begin{align*}
 \sup_{X\in V (\mu,\sigma)} \mathrm{TCRTE}_{\alpha,p}(X)&=\frac{\sigma\sqrt{d}}{|\alpha-1|(1-p)},~\alpha>\frac{1}{2},
 \end{align*}
which can be obtained by the
worst-case distribution of r.v. $X_{\ast}$ characterized by
\begin{align*}
F_{X_{\ast}}^{-1}(u)=
\begin{cases}
\mu+\sigma\frac{\mathrm{sign}(\alpha-1)}{\sqrt{d}}\left[1-\alpha\left(\frac{u-p}{1-p}\right)^{\alpha-1}\right],~u\in[u_{0},1],\\
\mu+\sigma\frac{|\alpha-1|(1-p)b}{\sqrt{d}},~u\in[0,u_{0}),
\end{cases}
\end{align*}
where $d=\frac{1-u_{0}}{u_{0}}-\frac{2(1-u_{0})^{\alpha}}{u_{0}(1-p)^{\alpha-1}}+\frac{(1-u_{0})^{2\alpha-1}}{(1-p)^{2\alpha-2}}\left[\frac{1}{u_{0}}+\frac{(\alpha-1)^{2}}{2\alpha-1}\right]$.

Letting $\alpha=2$ in Eq. (\ref{a12}), we obtain new type tail Gini functional, denoted by $\mathrm{TNGini}_{p}(X)$, for instance, see Zuo and Yin (2023), i.e.,
\begin{align*}
\mathrm{TNGini}_{p}(X)
=\int_{x_{p}}^{+\infty}\left[\frac{\overline{F}_{X}(x)}{1-p}-\left(\frac{\overline{F}_{X}(x)}{1-p}\right)^{2}\right]\mathrm{d}x.
\end{align*}
In this case, $\hat{g}_{\alpha,p}(u)=\hat{g}_{p}(u)=-\left[\frac{1-u}{1-p}-\left(\frac{1-u}{1-p}\right)^{2}\right]\mathbb{I}_{[p,1]}(u)$, $u\in[0,1]$, and
\begin{align*}
 [\hat{g}_{p}(u)]_{\ast}=
 \begin{cases}
 b u,~u\in[0,\sqrt{p}],\\
 -\left[\frac{1-u}{1-p}-\left(\frac{1-u}{1-p}\right)^{2}\right],~u\in(\sqrt{p},1],
 \end{cases}
 \end{align*}
 where $b=\frac{-1}{\sqrt{p}}\left[\frac{1}{1+\sqrt{p}}-\left(\frac{1}{1+\sqrt{p}}\right)^{2}\right]$.
 Hence, we get
\begin{align*}
 \sup_{X\in V (\mu,\sigma)} \mathrm{TNGini}_{p}(X)&=\frac{\sigma\sqrt{d}}{(1-p)},
 \end{align*}
which can be obtained by the
worst-case distribution of r.v. $X_{\ast}$ characterized by
\begin{align*}
F_{X_{\ast}}^{-1}(u)=
\begin{cases}
\mu+\sigma\frac{1}{\sqrt{d}}\left[1-2\left(\frac{u-p}{1-p}\right)\right],~u\in[\sqrt{p},1],\\
\mu+\sigma\frac{(1-p)b}{\sqrt{d}},~u\in[0,\sqrt{p}),
\end{cases}
\end{align*}
where $d=\frac{1-\sqrt{p}}{\sqrt{p}}-\frac{2(1-\sqrt{p})^{2}}{\sqrt{p}(1-p)}+\frac{(1-\sqrt{p})^{3}}{(1-p)^{2}}\left[\frac{1}{\sqrt{p}}+\frac{1}{3}\right]$.

Note that when $\alpha\rightarrow 1$, tail-based cumulative residual Tsallis entropy of order $\alpha$ will be tail-based cumulative residual entropy as following Example \ref{ex.7}.
\begin{example}\label{ex.7}A random variable $X$, with distribution function $F_{X}(x)$ and tail distribution function $\overline{F}_{X}(x)$, has its the tail-based cumulative residual entropy,  denoted by $\mathrm{TCRE}_{p}(X)$,  as (see Hu and Chen (2020))
 \begin{align*}
 \mathrm{TCRE}_{p}(X)
 &=-\int_{x_{p}}^{+\infty}\frac{\overline{F}_{X}(x)}{1-p}\log\left(\frac{\overline{F}_{X}(x)}{1-p}\right)\mathrm{d}x.
\end{align*}
In this case, $\psi(x)=1$, $t=x_{p}$ and
$g(u)=-u\log(u)$, $u\in[0,1]$, in Corollary \ref{co.2}. Since $\hat{g}_{t}(u)=\hat{g}_{p}(u)=\left(\frac{1-u}{1-p}\right)\log\left(\frac{1-u}{1-p}\right)\mathbb{I}_{[p,1]}(u)$ is not convex in $[0,1]$,
\begin{align*}
 [\hat{g}_{p}(u)]_{\ast}=
 \begin{cases}
 b u,~u\in[0,u_{0}],\\
 \left(\frac{1-u}{1-p}\right)\log\left(\frac{1-u}{1-p}\right),~u\in(u_{0},1],
 \end{cases}
 \end{align*}
 where $b=\frac{1-u_{0}}{(1-p)u_{0}}\log\left(\frac{1-u_{0}}{1-p}\right)$, and $u_{0}\in[p,1]$ is the solution to the following equation: $u_{0}+\log\left(\frac{1-u_{0}}{1-p}\right)=0$. For instance, $p=0.9$, then $u_{0}\approx0.96178$, $\hat{g}_{p}(u)~ \mathrm{and} ~[\hat{g}_{p}(u)]_{\ast}$ are plotted in Fig. 4(c).
 So that
 \begin{align*}
 \sup_{X\in V (\mu,\sigma)} \mathrm{TCRE}_{p}(X)&=\frac{\sigma\sqrt{d}}{1-p},
 \end{align*}
 which can be obtained by the
worst-case distribution of r.v. $X_{\ast}$ characterized by
\begin{align*}
F_{X_{\ast}}^{-1}(u)=
\begin{cases}
\mu-\frac{\sigma}{\sqrt{d}}\left[\log\left(\frac{1-u}{1-p}\right)+1\right],~u\in[u_{0},1],\\
\mu+\sigma\frac{b(1-p)}{\sqrt{d}},~u\in[0,u_{0}),
\end{cases}
\end{align*}
where $d=\frac{1-u_{0}}{u_{0}}\left(\log\left(\frac{1-u_{0}}{1-p}\right)\right)^{2}+(1-u_{0})$.
\end{example}

\begin{example}\label{ex.8} The new type tail-based extended Gini coefficient of a random variable $X$, with distribution function $F_{X}(x)$ and tail distribution function $\overline{F}_{X}(x)$, denoted by $\mathrm{TNEGini}_{r,p}(X)$, given by
 \begin{align*}
 \mathrm{TNEGini}_{r,p}(X)=-\int_{x_{p}}^{+\infty}2\left[\frac{F_{X}(x)-p}{1-p}+\left(\frac{\overline{F}_{X}(x)}{1-p}\right)^{r}-1\right]\mathrm{d}x.
 \end{align*}
 In this case, $\psi(x)=1$,  $t=x_{p}$ and
$g(u)=-2(1-u+u^{r}-1)$, $r>1$, $u\in[0,1]$, in Corollary \ref{co.2}. Since $\hat{g}_{t}(u)=\hat{g}_{r,p}(u)=2\left[\frac{u-p}{1-p}+\left(\frac{1-u}{1-p}\right)^{r}-1\right]\mathbb{I}_{[p,1]}(u)$ is not convex in $[0,1]$,
 \begin{align*}
 [\hat{g}_{r,p}(u)]_{\ast}=
 \begin{cases}
 b u,~u\in[0,u_{0}],\\
 2\left[\frac{u-p}{1-p}+\left(\frac{1-u}{1-p}\right)^{r}-1\right],~u\in(u_{0},1],
 \end{cases}
 \end{align*}
 where $b=\frac{2}{u_{0}}\left[\frac{u_{0}-p}{1-p}+\left(\frac{1-u_{0}}{1-p}\right)^{r}-1\right]$, and $u_{0}\in[p,1]$ is the solution to the following equation: $(1-u_{0})^{r}+r u_{0}(1-u_{0})^{r-1}-(1-p)^{r-1}=0$. For instance, $r=2$ and $p=0.5$, then $u_{0}=\sqrt{0.5}$, $\hat{g}_{r,p}(u)~ \mathrm{and} ~[\hat{g}_{r,p}(u)]_{\ast}$ are plotted in Fig. 5(a).
 So that
 \begin{align*}
 \sup_{X\in V (\mu,\sigma)} \mathrm{TNEGini}_{r,p}(X)&=\frac{2\sqrt{d}}{(1-p)}\sigma,
 \end{align*}
which can be obtained by the
worst-case distribution of r.v. $X_{\ast}$ characterized by
\begin{align*}
F_{X_{\ast}}^{-1}(u)=
\begin{cases}
\mu+\frac{\sigma}{\sqrt{d}}\left[1-r\left(\frac{1-u}{1-p}\right)^{r-1}\right],~u\in[u_{0},1],\\
\mu+\sigma\frac{b(1-p)}{2\sqrt{d}},~u\in[0,u_{0}),
\end{cases}
\end{align*}
with $d=\frac{1-u_{0}}{u_{0}}+\frac{(1-u_{0})^{2r-1}}{(1-p)^{2r-2}}\left(\frac{1}{u_{0}}-\frac{(r-1)^{2}}{2r-1}\right)-\frac{2(1-u_{0})^{r}}{u_{0}(1-p)^{r-1}}$.
\end{example}

\begin{remark}\label{re.6}
Note that $\psi(x)=1$, $t=x_{p}$ and $g(u)=-2(1-p)^{r-2}(u^{r}-u)$, $r>1,~u\in[0,1]$, in Corollary \ref{co.2}, we can obtain the tail-based extended Gini coefficient ($\mathrm{TEGini}_{r,p}(X)$), see, e.g., Berkhouch et al. (2018). In this case, $\hat{g}_{t}(u)=\hat{h}_{r,p}(u)=\frac{2}{(1-p)^{2}}[(1-u)^{r}+(1-p)^{r-1}u-(1-p)^{r-1}]\mathbb{I}_{[p,1]}(u)$. We find that $\hat{h}_{r,p}(u)=(1-p)^{r-2}\hat{g}_{r,p}(u)$, then,
\begin{align*}
 \sup_{X\in V (\mu,\sigma)} \mathrm{TEGini}_{r,p}(X)&=2(1-p)^{r-3}\sqrt{d}\sigma,
 \end{align*}
which can be obtained by the
worst-case distribution of r.v. $X_{\ast}$ characterized by
\begin{align*}
F_{X_{\ast}}^{-1}(u)=
\begin{cases}
\mu+\frac{\sigma}{\sqrt{d}}\left[1-r\left(\frac{1-u}{1-p}\right)^{r-1}\right],~u\in[u_{0},1],\\
\mu+\frac{b(1-p)\sigma}{2\sqrt{d}},~u\in[0,u_{0}),
\end{cases}
\end{align*}
where $b$, $d$ and $u_{0}$ are the same as those in Example \ref{ex.8}.
\end{remark}
When $r=2$, $\mathrm{TNEGini}_{r,p}$ will reduce to tail Gini functional ($\mathrm{TGini}_{p}(X)$), see, e.g., Furman et al. (2017),
\begin{align*}
 \mathrm{TGini}_{p}(X)=-\int_{x_{p}}^{+\infty}2\left[\frac{F_{X}(x)-p}{1-p}+\left(\frac{\overline{F}_{X}(x)}{1-p}\right)^{2}-1\right]\mathrm{d}x.
 \end{align*}
 In this case, $g(u)=-2(-u+u^{2}),~u\in[0,1]$. Since $\hat{g}_{r,p}(u)=\hat{g}_{p}(u)=2\left[\frac{u-p}{1-p}+\left(\frac{1-u}{1-p}\right)^{2}-1\right]\mathbb{I}_{[p,1]}(u)$ is not convex in $[0,1]$,
 \begin{align*}
 [\hat{g}_{p}(u)]_{\ast}=
 \begin{cases}
 b u,~u\in[0,\sqrt{p}],\\
 2\left[\frac{u-p}{1-p}+\left(\frac{1-u}{1-p}\right)^{2}-1\right],~u\in(\sqrt{p},1],
 \end{cases}
 \end{align*}
 where $b=\frac{2}{\sqrt{p}}\left[\frac{\sqrt{p}-p}{1-p}+\left(\frac{1-\sqrt{p}}{1-p}\right)^{2}-1\right]$.
Hence, we obtain
\begin{align*}
 \sup_{X\in V (\mu,\sigma)} \mathrm{TGini}_{p}&=\frac{2\sqrt{d}}{(1-p)}\sigma,
 \end{align*}
which can be obtained by the
worst-case distribution of r.v. $X_{\ast}$ characterized by
\begin{align*}
F_{X_{\ast}}^{-1}(u)=
\begin{cases}
\mu+\frac{\sigma}{\sqrt{d}}\left[1-2\left(\frac{1-u}{1-p}\right)\right],~u\in[\sqrt{p},1],\\
\mu+\sigma\frac{b(1-p)}{2\sqrt{d}},~u\in[0,\sqrt{p}),
\end{cases}
\end{align*}
with $d=\frac{1-\sqrt{p}}{\sqrt{p}}+\frac{(1-\sqrt{p})^{3}}{(1-p)^{2}}\left(\frac{1}{\sqrt{p}}-\frac{1}{3}\right)-\frac{2(1-\sqrt{p})^{2}}{\sqrt{p}(1-p)}$.

%%%%%%%%%%%%%%%%%%%%%%%%%%

\begin{example}\label{ex.9}A random variable $X$, with distribution function $F_{X}(x)$,  has its
 the dynamic cumulative Tsallis entropy,
  denoted by $\mathrm{DCT}_{\alpha,(t)}(X)$ (for $X\in L^{0}_{+}$, see Cal\`{\i} et al. (2017)), as
\begin{align*}
\mathrm{DCT}_{\alpha,(t)}(X)
&=\int_{-\infty}^{t}\frac{1}{\alpha-1}\left[\frac{F_{X}(x)}{F_{X}(t)}-\left(\frac{F_{X}(x)}{F_{X}(t)}\right)^{\alpha}\right]\mathrm{d}x.
\end{align*}
In this case, $\psi(x)=1$ and $g(u)=\frac{1}{\alpha-1}\left[(1-u)-(1-u)^{\alpha}\right],~\alpha>0,~\alpha\neq1,~u\in[0,1]$ in Corollary \ref{co.3}. Since \\ $\hat{g}_{(t)}(u)=\hat{g}_{\alpha,(t)}(u)=-\frac{1}{\alpha-1}\left[\frac{u}{F_{X}(t)}-\left(\frac{u}{F_{X}(t)}\right)^{\alpha}\right]\mathbb{I}_{[0,F_{X}(t)]}(u)$ is not convex in $[0,1]$,
\begin{align*}
 [\hat{g}_{\alpha,(t)}(u)]_{\ast}=
 \begin{cases}
 b_{1} u-b_{1},~u\in(u_{1},1],\\
 -\frac{1}{\alpha-1}\left[\frac{u}{F_{X}(t)}-\left(\frac{u}{F_{X}(t)}\right)^{\alpha}\right],~u\in[0,u_{1}],
 \end{cases}
 \end{align*}
 where $b_{1}=\frac{1}{(\alpha-1)(1-u_{1})}\left[\frac{u_{1}}{F_{X}(t)}-\left(\frac{u_{1}}{F_{X}(t)}\right)^{\alpha}\right]$, and $u_{1}\in[0,F_{X}(t)]$ is the solution to the following equation: $(F_{X}(t))^{\alpha-1}-u_{1}^{\alpha-1}[u_{1}+\alpha(1-u_{1})]=0$. For instance, $\alpha=\frac{2}{3}$ and $F_{X}(t)=0.2$, then $u_{1}\approx0.06525$, $\hat{g}_{\alpha,(t)}(u)~ \mathrm{and} ~[\hat{g}_{\alpha,(t)}(u)]_{\ast}$ are plotted in Fig. 5(b).
 Therefore, we get
 \begin{align*}
 \sup_{X\in V (\mu,\sigma)} \mathrm{DCT}_{\alpha,(t)}(X)&=\frac{\sigma\sqrt{d_{1}}}{|\alpha-1|F_{X}(t)},~\alpha>\frac{1}{2},
 \end{align*}
which can be obtained by the
worst-case distribution of r.v. $X_{\ast}$ characterized by
\begin{align*}
F_{X_{\ast}}^{-1}(u)=
\begin{cases}
\mu-\sigma\frac{\mathrm{sign}(\alpha-1)}{\sqrt{d_{1}}}\left[1-\alpha\left(\frac{u}{F_{X}(t)}\right)^{\alpha-1}\right],~u\in[0,u_{1}],\\
\mu+\sigma\frac{|\alpha-1|F_{X}(t)b_{1}}{\sqrt{d_{1}}},~u\in(u_{1},1],
\end{cases}
\end{align*}
where $d_{1}=\frac{u_{1}}{1-u_{1}}-\frac{2(u_{1})^{\alpha}}{(1-u_{1})(F_{X}(t))^{\alpha-1}}+\frac{(u_{1})^{2\alpha-1}}{(F_{X}(t))^{2\alpha-2}}\left[\frac{u_{1}}{1-u_{1}}+\frac{\alpha^{2}}{2\alpha-1}\right]$.
\end{example}
When $\alpha=2$, we call $\mathrm{DCT}_{\alpha,(t)}(X)$ as dynamic Gini functional ($\mathrm{DGini}_{(t)}(X)$):
\begin{align*}
\mathrm{DGini}_{(t)}(X)
&=\int_{-\infty}^{t}\left[\frac{F_{X}(x)}{F_{X}(t)}-\left(\frac{F_{X}(x)}{F_{X}(t)}\right)^{2}\right]\mathrm{d}x.
\end{align*}
In this case, $g(u)=\left[\frac{1-u}{F_{X}(t)}-\left(\frac{1-u}{F_{X}(t)}\right)^{2}\right],~u\in[0,1]$, $\hat{g}_{(t)}(u)=-\left[\frac{u}{F_{X}(t)}-\left(\frac{u}{F_{X}(t)}\right)^{2}\right]\mathbb{I}_{[0,F_{X}(t)]}(u)$, and
\begin{align*}
 [\hat{g}_{(t)}(u)]_{\ast}=
 \begin{cases}
 b_{1} u-b_{1},~u\in(1-\sqrt{1-F_{X}(t)},1],\\
 -\left[\frac{u}{F_{X}(t)}-\left(\frac{u}{F_{X}(t)}\right)^{2}\right],~u\in[0,1-\sqrt{1-F_{X}(t)}],
 \end{cases}
 \end{align*}
 where $b_{1}=\frac{1}{\sqrt{1-F_{X}(t)}}\left[\frac{1-\sqrt{1-F_{X}(t)}}{F_{X}(t)}-\left(\frac{1-\sqrt{1-F_{X}(t)}}{F_{X}(t)}\right)^{2}\right]$.
Thus, we have
\begin{align*}
 \sup_{X\in V (\mu,\sigma)} \mathrm{DGini}_{(t)}(X)&=\frac{\sigma\sqrt{d_{1}}}{F_{X}(t)},
 \end{align*}
 which can be obtained by the
worst-case distribution of r.v. $X_{\ast}$ characterized by
\begin{align*}
F_{X_{\ast}}^{-1}(u)=
\begin{cases}
\mu-\sigma\frac{1}{\sqrt{d_{1}}}\left[1-2\left(\frac{u}{F_{X}(t)}\right)\right],~u\in[0,1-\sqrt{1-F_{X}(t)}],\\
\mu+\sigma\frac{F_{X}(t)b_{1}}{\sqrt{d_{1}}},~u\in(1-\sqrt{1-F_{X}(t)},1],
\end{cases}
\end{align*}
where $d_{1}=\frac{1-\sqrt{1-F_{X}(t)}}{\sqrt{1-F_{X}(t)}}-\frac{2\left(1-\sqrt{1-F_{X}(t)}\right)^{2}}{(\sqrt{1-F_{X}(t)})(F_{X}(t))}+\frac{\left(1-\sqrt{1-F_{X}(t)}\right)^{3}}{(F_{X}(t))^{2}}\left[\frac{1-\sqrt{1-F_{X}(t)}}{\sqrt{1-F_{X}(t)}}+\frac{4}{3}\right]$.

Note that when $\alpha\rightarrow 1$, dynamic cumulative Tsallis entropy of order $\alpha$ will be dynamic cumulative past entropy as following Example \ref{ex.10}.
\begin{example}\label{ex.10}The dynamic cumulative past entropy of a random variable $X$, with distribution function $F_{X}(x)$, denoted by $D\mathcal{CE}_{(t)}(X)$, see, e.g., Navarro et al. (2010), given by
\begin{align*}
 D\mathcal{CE}_{(t)}(X)
 &=-\int_{0}^{t}\frac{F_{X}(x)}{F_{X}(t)}\log\left(\frac{F_{X}(x)}{F_{X}(t)}\right)\mathrm{d}x.
 \end{align*}
 In this case, $\psi(x)=1$ and
$g(u)=-(1-u)\log\left(1-u\right),~u\in[0,1]$, in Corollary \ref{co.3}. Since $\hat{g}_{(t)}(u)=\frac{u}{F_{X}(t)}\log\left(\frac{u}{F_{X}(t)}\right)\mathbb{I}_{[0,F_{X}(t)]}(u)$ is not convex in $[0,1]$,
\begin{align*}
 [\hat{g}_{(t)}(u)]_{\ast}=
 \begin{cases}
 b_{1} u-b_{1},~u\in(u_{1},1],\\
 \frac{u}{F_{X}(t)}\log\left(\frac{u}{F_{X}(t)}\right),~u\in[0,u_{1}],
 \end{cases}
 \end{align*}
 $b_{1}=\frac{u_{1}}{(u_{1}-1)F_{X}(t)}\log\left(\frac{u_{1}}{F_{X}(t)}\right)$, and
 $u_{1}\in[0,F_{X}(t)]$ is the solution to the following equation: $u_{1}-1-\log\left(\frac{u_{1}}{F_{X}(t)}\right)=0$. For instance, $F_{X}(t)=0.9$, then $u_{1}\approx0.60834$, $\hat{g}_{(t)}(u)~ \mathrm{and} ~[\hat{g}_{(t)}(u)]_{\ast}$ are plotted in Fig. 5(c).
  Hence, we get
\begin{align*}
\sup_{X\in V (\mu,\sigma)} D\mathcal{CE}(X)&=\frac{\sigma\sqrt{d_{1}}}{F_{X}(t)},
\end{align*}
 which can be obtained by the
worst-case distribution of r.v. $X_{\ast}$ characterized by
\begin{align*}
F_{X_{\ast}}^{-1}(u)=
\begin{cases}
\mu+\frac{\sigma}{\sqrt{d_{1}}}\left[\log \left(\frac{u}{F_{X}(t)}\right)+1\right],~u\in[0,u_{1}],\\
\mu+\sigma\frac{b_{1}F_{X}(t)}{\sqrt{d_{1}}},~u\in(u_{1},1],
\end{cases}
\end{align*}
where $d_{1}=u_{1}+\frac{u_{1}(2u_{1}-1)}{u_{1}-1}\left(\log\left(\frac{u_{1}}{F_{X}(t)}\right)\right)^{2}$.
\end{example}
\section{Applications to weighted entropies\label{sec:5}}
In this section, we apply the results of the previous section to  some the commonly used weighted entropies, including (dynamic) weighted Gini, (dynamic) weighted CRE, (dynamic) weighted CT and (dynamic) weighted EGini.

 \begin{example}\label{ex.11}The the weighted cumulative Tsallis entropy of order $\alpha$ of a random variable $X$, with distribution function $F_{X}(x)$, denoted by $\mathrm{WCT}_{\alpha}(X)$ (for $X\in L_{+}^{0}$, see, e.g., Chakraborty and Pradhan (2023)), given by
\begin{align*}
\mathrm{WCT}_{\alpha}(X)
=\int_{-\infty}^{+\infty}\frac{1}{\alpha-1}x\left[F_{X}(x)-(F_{X}(x))^{\alpha}\right]\mathrm{d}x.
\end{align*}
In this case, $\psi(x)=x$ and $g(u)=\frac{1}{\alpha-1}(1-u-(1-u)^{\alpha}), ~\alpha>0,~\alpha\neq1,~u\in[0,1]$ in Theorem \ref{th.2}. Since $\hat{g}(u)=-g(1-u)$ is convex in $[0,1]$, $\hat{g}_{\ast}(u)=-g(1-u)$.
Thus,
\begin{align*}
 \sup_{X\in V (\mu_{\Psi},\sigma_{\Psi})} \mathrm{WCT}_{\alpha}(X)&=\frac{\sqrt{\mathrm{Var}(X^{2})}}{2\sqrt{2\alpha-1}},~\alpha>\frac{1}{2},
\end{align*}
which can be obtained by the
worst-case distribution of r.v. $X_{\ast}$ characterized by
$$(F_{X_{\ast}}^{-1}(u))^{2}=\mathrm{E}[X^{2}]+\frac{1}{2}\sqrt{\mathrm{Var}(X^{2})}\frac{\sqrt{2\alpha-1}}{\alpha-1}\left(\alpha u^{\alpha-1}-1\right).$$
\end{example}

When $\alpha=2$, $g(u)=u(1-u)$, we call it as weighted Gini functional, denoted by $\mathrm{W\mathcal{G}ini}(X)$, i.e.,
\begin{align*}
\mathrm{W\mathcal{G}ini}(X)
=\int_{-\infty}^{+\infty}xF_{X}(x)\overline{F}_{X}(x)\mathrm{d}x.
\end{align*}
Hence,
\begin{align*}
\sup_{X\in V (\mu_{\Psi},\sigma_{\Psi})} \mathrm{W\mathcal{G}ini}(X)=\frac{\sqrt{\mathrm{Var}(X^{2})}}{2\sqrt{3}},
\end{align*}
which can be obtained by the
worst-case distribution of r.v. $X_{\ast}$ characterized by
$$(F_{X_{\ast}}^{-1}(u))^{2}=\mathrm{E}[X^{2}]+\frac{\sqrt{3}}{2}\sqrt{\mathrm{Var}(X^{2})}\left(2 u-1\right).$$
\begin{example}\label{ex.12}A random variable $X$, with distribution function $F_{X}(x)$ and tail distribution function $\overline{F}_{X}(x)$,  has its
 the  weighted cumulative residual Tsallis entropy of order $\alpha$,
  denoted by $\mathrm{WCRT}_{\alpha}(X)$ (for $X\in L_{+}^{0}$, see Chakraborty and Pradhan (2023)), as
 \begin{align*}
\mathrm{WCRT}_{\alpha}(X)
&=\int_{-\infty}^{+\infty}\frac{1}{\alpha-1}x\left[\overline{F}_{X}(x)-(\overline{F}_{X}(x))^{\alpha}\right]\mathrm{d}x.
\end{align*}
In this case, $\psi(x)=x$ and $g(u)=\frac{1}{\alpha-1}[u-u^{\alpha}], ~\alpha>0,~\alpha\neq1,~u\in[0,1]$ in Theorem \ref{th.2}. Since $\hat{g}(u)=-g(1-u)$ is convex in $[0,1]$, $\hat{g}_{\ast}(u)=-g(1-u)$.
Thus,
\begin{align*}
\sup_{X\in V (\mu_{\Psi},\sigma_{\Psi})} \mathrm{WCRT}_{\alpha}(X)&=\frac{\sqrt{\mathrm{Var}(X^{2})}}{2\sqrt{2\alpha-1}},~\alpha>\frac{1}{2},
\end{align*}
which can be obtained by the
worst-case distribution of r.v. $X_{\ast}$ characterized by
$$(F_{X_{\ast}}^{-1}(u))^{2}=\mathrm{E}[X^{2}]+\frac{1}{2}\sqrt{\mathrm{Var}(X^{2})}\frac{\sqrt{2\alpha-1}}{\alpha-1}\left[1-\alpha(1-u)^{\alpha-1}\right].$$
\end{example}

 \begin{example}\label{ex.13} The weighted cumulative residual entropy  with weight function $\psi$ of a random variable $X$, with tail distribution function $\overline{F}_{X}(x)$, denoted by $\mathrm{WGCRE}_{\psi}(X)$ ( see, e.g., Suhov and Yasaei Sekeh (2015)), given by
\begin{align*}
 \mathrm{WGCRE}_{\psi}(X)
 &=\int_{0}^{+\infty}\psi(x)\overline{F}_{X}(x)\left[-\log\left(\overline{F}_{X}(x)\right)\right]\mathrm{d}x.
 \end{align*}
 In this case, $g(u)=-u\mathrm{log}(u),~u\in[0,1]$ in Theorem \ref{th.2}. Because $\hat{g}(u)=-g(1-u)$ is convex in $[0,1]$, $\hat{g}_{\ast}(u)=-g(1-u)$.
Thus,
\begin{align*}
 \sup_{X\in V (\mu_{\Psi},\sigma_{\Psi})} \mathrm{WGCRE}_{\psi}(X)=\sigma_{\Psi},
\end{align*}
which can be obtained by the
worst-case distribution of r.v. $X_{\ast}$ characterized by
$$\Psi(F_{X_{\ast}}^{-1}(u))=\mu_{\Psi}-\sigma_{\Psi}(\log (1-u)+1).$$
 \end{example}

 When $\psi(x)= x$, $\mathrm{WGCRE}_{\psi}(X)$ will be weighted cumulative residual entropy ($\mathcal{E}^{w}(X)$), see, e.g., Misagh et al. (2011) and  Mirali et al. (2017),
\begin{align*}
\mathcal{E}^{w}(X)
&=-\int_{0}^{+\infty}x\overline{F}_{X}(x)\mathrm{log}(\overline{F}_{X}(x))\mathrm{d}x.
\end{align*}
Hence,
\begin{align*}
\sup_{X\in V (\mu_{\Psi},\sigma_{\Psi})}\mathcal{E}^{w}(X)=\frac{1}{2}\sqrt{\mathrm{Var}(X^{2})},
\end{align*}
which can be obtained by the
worst-case distribution of r.v. $X_{\ast}$ characterized by
$$(F_{X_{\ast}}^{-1}(u))^{2}=\mathrm{E}[X^{2}]-\frac{1}{2}\sqrt{\mathrm{Var}(X^{2})}(\log (1-u)+1).$$

 \begin{example}\label{ex.14} A random variable $X$, with distribution function $F_{X}(x)$,  has its
 weighted cumulative entropy with weight function $\psi$,
  denoted by $\mathrm{WGCE}_{\psi}(X)$ (see, e.g., Suhov and Yasaei Sekeh (2015)), as
\begin{align*}
\mathrm{ WGCE}_{\psi}(X)
 &=\int_{0}^{+\infty}\psi(x)F_{X}(x)\left[-\log\left(F_{X}(x)\right)\right]\mathrm{d}x.
\end{align*}
In this case, $g(u)=(1-u)[-\log(1-u)]$ in Theorem \ref{th.2}.  Since $\hat{g}(u)=-g(1-u)$ is convex in $[0,1]$, $\hat{g}_{\ast}(u)=-g(1-u)$.
Thus,
\begin{align*}
 \sup_{X\in V (\mu_{\Psi},\sigma_{\Psi})} \mathrm{WGCE}_{\psi}(X)=\sigma_{\Psi},
\end{align*}
which can be obtained by the
worst-case distribution of r.v. $X_{\ast}$ characterized by
$$\Psi(F_{X_{\ast}}^{-1}(u))=\mu_{\Psi}+\sigma_{\Psi}(\log u+1).$$
\end{example}
 When $\psi(x)= x$, $\mathrm{WGCE}_{\psi}(X)$ will be weighted cumulative entropy ($\mathcal{CE}^{w}(X)$), see, e.g., Misagh et al. (2011) and Mirali and Baratpour (2017a),
\begin{align*}
\mathcal{CE}^{w}(X)&=-\int_{0}^{+\infty}xF_{X}(x)\mathrm{log}(F_{X}(x))\mathrm{d}x.
\end{align*}
Hence,
\begin{align*}
\sup_{X\in V (\mu_{\Psi},\sigma_{\Psi})}\mathcal{CE}^{w}(X)=\frac{1}{2}\sqrt{\mathrm{Var}(X^{2})},
\end{align*}
which can be obtained by the
worst-case distribution of r.v. $X_{\ast}$ characterized by
$$(F_{X_{\ast}}^{-1}(u))^{2}=\mathrm{E}[X^{2}]+\frac{1}{2}\sqrt{\mathrm{Var}(X^{2})}(\log u+1).$$

%%%%%%%%%%%%%%%%%%%%%%%%%%%%%%%%%%%%%%%%
%%%%%%%%%%%%%%%%%%%%%%%%%%%%%%%%%%%%%%%%
\begin{example}\label{ex.15} The generalized dynamic weighted cumulative residual entropy with weight function $\psi$ of a random variable $X$, with tail distribution function $\overline{F}_{X}(x)$, denoted by $\mathrm{DWGCRE}_{\psi,t}(X)$ (see, e.g., Miralia and Baratpour (2017b)), given by
\begin{align*}
 \mathrm{DWGCRE}_{\psi,t}(X)
 &=\int_{t}^{+\infty}\psi(x)\overline{F}_{X}(x)\left[-\log\left(\overline{F}_{X}(x)\right)\right]\mathrm{d}x.
\end{align*}
In this case, $g(u)=-u\log(u)$, $u\in[0,1]$ in Corollary \ref{co.2}. Because  $\hat{g}_{t}(u)=\left(\frac{1-u}{1-F_{X}(t)}\right)\left[\log\left(\frac{1-u}{1-F_{X}(t)}\right)\right]\mathbb{I}_{[F_{X}(t),1]}(u)$ is not convex in $[0,1]$,
\begin{align*}
 [\hat{g}_{t}(u)]_{\ast}=
 \begin{cases}
 b u,~u\in[0,u_{0}],\\
 \left(\frac{1-u}{1-F_{X}(t)}\right)\log\left(\frac{1-u}{1-F_{X}(t)}\right),~u\in(u_{0},1],
 \end{cases}
 \end{align*}
 where $b=\frac{1-u_{0}}{(1-F_{X}(t))u_{0}}\log\left(\frac{1-u_{0}}{1-F_{X}(t)}\right)$, and $u_{0}\in[F_{X}(t),1]$ is the solution to the following equation: $u_{0}+\log\left(\frac{1-u_{0}}{1-F_{X}(t)}\right)=0$.
Thus,
\begin{align*}
  \sup_{X\in V (\mu_{\Psi},\sigma_{\Psi})} \mathrm{DWGCRE}_{\psi,t}(X)&=\frac{\sigma_{\Psi}\sqrt{d}}{1-F_{X}(t)},
 \end{align*}
 which can be obtained by the
worst-case distribution of r.v. $X_{\ast}$ characterized by
\begin{align*}
\Psi(F_{X_{\ast}}^{-1}(u))=
\begin{cases}
\mu_{\Psi}-\frac{\sigma_{\Psi}}{\sqrt{d}}\left[\log\left(\frac{1-u}{1-F_{X}(t)}\right)+1\right],~u\in[u_{0},1],\\
\mu_{\Psi}+\sigma_{\Psi}\frac{b(1-F_{X}(t))}{\sqrt{d}},~u\in[0,u_{0}),
\end{cases}
\end{align*}
where $d=\frac{1-u_{0}}{u_{0}}\left(\log\left(\frac{1-u_{0}}{1-F_{X}(t)}\right)\right)^{2}+(1-u_{0})$.
\end{example}

Particularly, when
 $\psi(x)=x$, the $\mathrm{DWGCRE}_{\psi,t}(X)$ will reduce to the dynamic weighted cumulative residual entropy ($\mathrm{DWCRE}_{t}(X)$) ($X\in L_{+}^{0}$, also see Miralia and Baratpour (2017b)),
\begin{align*}
 \mathrm{DWCRE}_{t}(X)
 &=\int_{t}^{+\infty}x\overline{F}_{X}(x)\left[-\log\left(\overline{F}_{X}(x)\right)\right]\mathrm{d}x.
\end{align*}
Hence,
\begin{align*}
  \sup_{X\in V (\mu_{\Psi},\sigma_{\Psi})} \mathrm{DWCRE}_{t}(X)&=\frac{\sqrt{\mathrm{Var}(X^{2})}\sqrt{d}}{1-F_{X}(t)},
 \end{align*}
 which can be obtained by the
worst-case distribution of r.v. $X_{\ast}$ characterized by
\begin{align*}
[F_{X_{\ast}}^{-1}(u)]^{2}=
\begin{cases}
\mathrm{E}(X^{2})-\frac{\sqrt{\mathrm{Var}(X^{2})}}{\sqrt{d}}\left[\log\left(\frac{1-u}{1-F_{X}(t)}\right)+1\right],~u\in[u_{0},1],\\
\mathrm{E}(X^{2})+\sqrt{\mathrm{Var}(X^{2})}\frac{b(1-F_{X}(t))}{\sqrt{d}},~u\in[0,u_{0}),
\end{cases}
\end{align*}
where $b$, $d$ and $u_{0}$ are the same as those in Example \ref{ex.15}.

\begin{example}\label{ex.16} A random variable $X$, with distribution function $F_{X}(x)$,  has its
 generalized dynamic weighted cumulative entropy with weight function $\psi$,
  denoted by $\mathrm{DWGCE}_{\psi,t}(X)$ ($X\in L_{+}^{0}$, see, e.g., Miralia and Baratpour (2017b)), as
\begin{align*}
 \mathrm{DWGCE}_{\psi,(t)}(X)
 &=\int_{0}^{t}\psi(x)\frac{F_{X}(x)}{F_{X}(t)}\left[-\log\left(\frac{F_{X}(x)}{F_{X}(t)}\right)\right]\mathrm{d}x.
\end{align*}
In this case, $g(u)=-(1-u)\log\left(1-u\right)$, $u\in[0,1]$ in Corollary \ref{co.3}. Since $\hat{g}_{(t)}(u)=\left(\frac{u}{F_{X}(t)}\right)\left[\log\left(\frac{u}{F_{X}(t)}\right)\right]\mathbb{I}_{[0,F_{X}(t)]}(u)$ is not convex in $[0,1]$,
 \begin{align*}
 [\hat{g}_{(t)}(u)]_{\ast}=
 \begin{cases}
 b_{1} u-b_{1},~u\in(u_{1},1],\\
 \frac{u}{F_{X}(t)}\log\left(\frac{u}{F_{X}(t)}\right),~u\in[0,u_{1}],
 \end{cases}
 \end{align*}
 $b_{1}=\frac{u_{1}}{(u_{1}-1)F_{X}(t)}\log\left(\frac{u_{1}}{F_{X}(t)}\right)$, and
 $u_{1}\in[0,F_{X}(t)]$ is the solution to the following equation: $u_{1}-1-\log\left(\frac{u_{1}}{F_{X}(t)}\right)=0$.
Thus,
\begin{align*}
\sup_{X\in V (\mu_{\Psi},\sigma_{\Psi})} \mathrm{DWGCE}_{\psi,(t)}(X)&=\frac{\sigma_{\Psi}\sqrt{d_{1}}}{F_{X}(t)},
\end{align*}
 which can be obtained by the
worst-case distribution of r.v. $X_{\ast}$ characterized by
\begin{align*}
\Psi(F_{X_{\ast}}^{-1}(u))=
\begin{cases}
\mu_{\Psi}+\frac{\sigma_{\Psi}}{\sqrt{d_{1}}}\left[\log \left(\frac{u}{F_{X}(t)}\right)+1\right],~u\in[0,u_{1}],\\
\mu_{\Psi}+\sigma_{\Psi}\frac{b_{1}F_{X}(t)}{\sqrt{d_{1}}},~u\in(u_{1},1],
\end{cases}
\end{align*}
where $d_{1}=u_{1}+\frac{u_{1}(2u_{1}-1)}{u_{1}-1}\left(\log\left(\frac{u_{1}}{F_{X}(t)}\right)\right)^{2}$.
\end{example}
In particular, when $\psi(x)=x$, the $\mathrm{DWGCE}_{\psi,(t)}(X)$ will reduce to the dynamic weighted cumulative entropy ($\mathrm{DWCE}_{(t)}(X)$) ($X\in L_{+}^{0}$, also see Miralia and Baratpour (2017b)),
\begin{align*}
 \mathrm{DWCE}_{(t)}(X)
 &=\int_{0}^{t}x\frac{F_{X}(x)}{F_{X}(t)}\left[-\log\left(\frac{F_{X}(x)}{F_{X}(t)}\right)\right]\mathrm{d}x.
\end{align*}
Hence,
\begin{align*}
\sup_{X\in V (\mu_{\Psi},\sigma_{\Psi})} \mathrm{DWCE}_{(t)}(X)&=\frac{\sqrt{\mathrm{Var}(X^{2})}\sqrt{d_{1}}}{F_{X}(t)},
\end{align*}
 which can be obtained by the
worst-case distribution of r.v. $X_{\ast}$ characterized by
\begin{align*}
[F_{X_{\ast}}^{-1}(u)]^{2}=
\begin{cases}
\mathrm{E}(X^{2})+\frac{\sqrt{\mathrm{Var}(X^{2})}}{\sqrt{d_{1}}}\left[\log \left(\frac{u}{F_{X}(t)}\right)+1\right],~u\in[0,u_{1}],\\
\mathrm{E}(X^{2})+\sqrt{\mathrm{Var}(X^{2})}\frac{b_{1}F_{X}(t)}{\sqrt{d_{1}}},~u\in(u_{1},1],
\end{cases}
\end{align*}
where $b_{1}$, $d_{1}$ and $u_{1}$ are the same as those in Example \ref{ex.16}.
\section{Applications to risk measures\label{sec:6}}
In this section, we apply the results of the previous section to some risk measures, including several  premium principles and  several shortfalls.
\subsection{Sharp upper bounds for premium principles\label{sec:6.1}}
 As we all known, the standard deviation and absolute deviation premium principles are defined as, respectively,
 \begin{align*}
 \mathrm{SD}(X)=\mu+\kappa\sqrt{\mathrm{Var}(X)},~\kappa>0,
 \end{align*}
 \begin{align*}
 \mathrm{AD}(X)=\mu+\kappa \mathrm{E}|X-\mu|,~\kappa>0,
 \end{align*}
 where $\mu$ and $\mathrm{Var}(X)$ denote the mean and variance of $X$, respectively. We know that $\mathrm{E}|X-\mu|\leq\sqrt{\mathrm{Var}(X)}$  is a direct consequence of Cauchy-Schwarz inequality. Therefore, we get $\mathrm{AD}(X) \leq \mathrm{SD}(X)$.\\
 Based entropy's premium principle is defined as
 \begin{align}\label{a13}
 \mathrm{B}\mathcal{EN}(X)=\mu+\kappa \mathcal{EN}(X),~\kappa>0,
 \end{align}
 where $\mathcal{EN}(X)$ is any entropy of $X$.

By Corollary \ref{co.1} and Eq. (\ref{a13}), we can provide the sharp upper bound of $\mathrm{B}\mathcal{EN}(X)$.
\begin{proposition}\label{pro.1}
Under conditions of Corollary \ref{co.1}, the sharp upper bound of based entropy premium principle is given by
\begin{align}\label{a14}
 \sup_{X\in V (\mu,\sigma)} \mathrm{B}\mathcal{EN}(X) =\mu+\kappa\sigma\sqrt{\int_{0}^{1}\left((-g(1-u))_{\ast}'\right)^{2}\mathrm{d}u},~\kappa>0,
\end{align}
if $(-g(1-u))_{\ast}' =0 ~(a.e.)$, then (\ref{a14}) can be obtained by any random variable $X \in
V (\mu, \sigma)$; If $(-g(1-u))_{\ast}' \neq0 ~(a.e.)$, then (\ref{a14})
 can be obtained by the
worst-case distribution of r.v. $X_{\ast}$ characterized by
$$F_{X_{\ast}}^{-1+}(u)=\mu+\sigma\frac{(-g(1-u))_{\ast}'}{\sqrt{\int_{0}^{1}\left((-g(1-u))_{\ast}'\right)^{2}\mathrm{d}u}}.$$
\end{proposition}
 From Proposition \ref{pro.1}, we can give sharp upper bounds of several commonly used premium principle.
 \\
 For $\kappa>0$, Gini's, cumulative entropy's, cumulative residual entropy's, cumulative Tsallis past entropy's ($\alpha>\frac{1}{2}$ and $\alpha\neq1$), cumulative residual Tsallis entropy's ($\alpha>\frac{1}{2}$ and $\alpha\neq1$) and extended Gini's ($r>1$)  premium principles are defined, respectively, as
 $$\mathrm{BGini}(X)=\mu+\kappa \mathrm{Gini}(X),$$
 $$\mathrm{B}\mathcal{CE}(X)=\mu+\kappa \mathcal{CE}(X),$$
  $$\mathrm{B}\mathcal{E}(X)=\mu+\kappa \mathcal{E}(X),$$
 $$\mathrm{BCT}_{\alpha}(X)=\mu+\kappa \mathrm{CT}_{\alpha}(X),$$
  $$\mathrm{BCRT}_{\alpha}(X)=\mu+\kappa \mathrm{CRT}_{\alpha}(X),$$
  $$\mathrm{BEGini}_{\alpha}(X)=\mu+\kappa \mathrm{EGini}_{r}(X),$$
  where $\mathrm{Gini}(X)$, $\mathcal{CE}(X)$, $\mathcal{E}(X)$, $\mathrm{CT}_{\alpha}(X)$, $\mathrm{CRT}_{\alpha}(X)$ and $\mathrm{EGini}_{r}(X)$ are defined in Section \ref{sec:4} (Examples \ref{ex.1}-\ref{ex.5}).

  Their sharp upper bounds are given by
 $$\sup_{X\in V (\mu,\sigma)} \mathrm{BGini}(X)=\mu+\kappa \frac{2\sigma}{\sqrt{3}},$$
 $$\sup_{X\in V (\mu,\sigma)} \mathrm{B}\mathcal{CE}(X)=\mu+\kappa \sigma,$$
  $$\sup_{X\in V (\mu,\sigma)} \mathrm{B}\mathcal{E}(X)=\mu+\kappa \sigma,$$
 $$\sup_{X\in V (\mu,\sigma)} \mathrm{BCT}_{\alpha}(X)=\mu+\kappa \frac{\sigma}{\sqrt{2\alpha-1}},~\alpha>\frac{1}{2},~\alpha\neq1,$$
  $$\sup_{X\in V (\mu,\sigma)} \mathrm{BCRT}_{\alpha}(X)=\mu+\kappa \frac{\sigma}{\sqrt{2\alpha-1}},~\alpha>\frac{1}{2},~\alpha\neq1,$$
  $$\sup_{X\in V (\mu,\sigma)} \mathrm{BEGini}_{\alpha}(X)=\mu+\kappa \frac{2(r-1)}{\sqrt{2r-1}}\sigma,~r>1.$$

  We observe that $\mathrm{AD}(X)\leq\sup_{X\in V (\mu,\sigma)} \mathrm{B}\mathcal{CE}(X)=\sup_{X\in V (\mu,\sigma)} \mathrm{B}\mathcal{E}(X)=\mathrm{SD}(X)\leq\sup \mathrm{BGini}(X)$ and $\sup_{X\in V (\mu,\sigma)} \mathrm{BCT}_{\alpha}(X)=\sup \mathrm{BCRT}_{\alpha}(X)$. Furthermore, the values of $\sup_{X\in V (\mu,\sigma)} \mathrm{BEGini}_{\alpha}(X)$, $\sup_{X\in V (\mu,\sigma)} \mathrm{BCT}_{\alpha}(X)$ and $\sup_{X\in V (\mu,\sigma)} \mathrm{BCRT}_{\alpha}(X)$ are very flexible, and we can choose different parameters to meet our needs.
\subsection{Worst-cases of shortfalls\label{sec:6.2}}
Furman et al. (2017) proposed the Gini shortfall (GS) to incorporate variability
in tail risk analysis, and GS at confidence level
$s \in(0, 1)$ for a r.v. $X$ with cdf $F_{V}$ is given by
\begin{align*}
\mathrm{GS}_{s}^{\tau} (X) = \mathrm{ES}_{s}(X) + \tau \mathrm{TGini}_{s}(X),
\end{align*}
where $\mathrm{ES}_{s}(X) = \frac{1}{1 - s}\int _{s}^{1}\mathrm{VaR}_{q}(X)\mathrm{d}q$ and $\mathrm{TGini}_{s}(X) = \frac{2}{(1 - s)^{2}}\int _{p}^{1}F_{X}^{-1}(v)(2v - (1 + s))\mathrm{d}v$ denote expected shortfall (ES) and  tail-Gini functional, respectively. In addition, $\mathrm{VaR}_{q}(X)=F_{X}^{-1}(q)$.

Let $\psi(x)=1$ and $t=x_{p}$, $p\in(0,1)$, in Corollary \ref{co.2}, we define tail-based entropy:
\begin{align}\label{a15}
  \mathcal{EN}_{p}(X)=  \int_{x_{p}}^{+\infty}g\left(\frac{\overline{F}_{X}(x)}{1-p}\right)\mathrm{d}x.
\end{align}

The risk measure of entropy-shortfall is defined as
\begin{align}\label{a16}
\mathrm{\mathcal{EN}S}_{p}^{\tau} (X) = \mathrm{ES}_{p}(X) + \tau \mathcal{EN}_{p}(X),~\tau\geq0.
\end{align}

Now, we give the worst-case of $\mathrm{\mathcal{EN}S}_{p}^{\tau} (X)$ as follows.
\begin{proposition}\label{pro.2}
Under conditions of Corollary \ref{co.1}, the worst-case of risk measure of entropy-shortfall is given by
\begin{align}\label{a17}
  \sup_{X\in V (\mu,\sigma)}\mathcal{EN}S_{p}^{\tau}(X)&=\mu+\sigma\sqrt{\int_{0}^{1}(h_{\tau,p\ast}'(u)-1)^{2}\mathrm{d}u},
\end{align}
where $h_{\tau,p}=\left[\frac{u-p}{1-p}-\tau g\left(\frac{1-u}{1-p}\right)\right]\mathbb{I}_{[p,1]}(u),~u\in[0,1],$ $\tau\geq0$,
if $h_{\tau,p\ast}'(u)=1~(a.e.)$, then (\ref{a17}) can be obtained by any random variable $X \in
V (\mu, \sigma)$; If $h_{\tau,p\ast}'(u)\neq1 ~(a.e.)$, then (\ref{a17})
 can be obtained by the
worst-case distribution of r.v. $X_{\ast}$ characterized by
$$F_{X_{\ast}}^{-1+}(u)=\mu+\sigma\frac{h_{\tau,p\ast}'(u)-1}{\sqrt{\int_{0}^{1}\left(h_{\tau,p\ast}'(u)-1\right)^{2}\mathrm{d}u}}.$$
\end{proposition}
\noindent $\mathbf{Proof.}$ Using Lemma \ref{le.2} and Eq. (\ref{a15}), we have
\begin{align*}
  \mathcal{EN}_{p}(X)= -\int_{0}^{1}F_{X}^{-1}(u)\mathrm{d}\left[g\left(\frac{1-u}{1-p}\right)\mathbb{I}_{[p,1]}(u)\right].
\end{align*}
Note that
\begin{align*}
\mathrm{ES}_{p}(X)=\int_{0}^{1}F_{X}^{-1}(u)\mathrm{d}\left[\frac{u-p}{1-p}\mathbb{I}_{[p,1]}(u)\right],
\end{align*}
combining with Eq. (\ref{a16}), thus, we get
\begin{align*}
\mathrm{\mathcal{EN}S}_{p}^{\tau} (X) = \int_{0}^{1}F_{X}^{-1}(u)\mathrm{d}h_{\tau,p}.
\end{align*}
Then, from Theorem \ref{th.1} we obtain desired results. $\hfill\square$

Next, some the commonly used shortfalls $\mathrm{GS}_{p}^{\tau} (X)$, $\mathrm{EGS}_{r,p}^{\tau} (X)$, $\mathrm{CRES}_{p}^{\tau} (X) $ and $\mathrm{CRTES}^{\tau}
_{\alpha,p}(X)$ are provided.

\begin{example}\label{ex.17} The Gini shortfall (GS) of a random variable $X$, with $p\in(0,1)$ and the loading
parameter $\tau \geq 0$, denoted by $\mathrm{GS}_{s}^{\tau} (X)$ (see Furman et al. (2017)), given by
 \begin{align*}
\mathrm{GS}_{p}^{\tau} (X) = \mathrm{ES}_{p}(X) + \tau \mathrm{TGini}_{p}(X),
\end{align*}
where $\mathrm{TGini}_{p}(X) = \frac{2}{(1 - p)^{2}}\int _{p}^{1}F_{X}^{-1}(v)(2v - (1 + p))\mathrm{d}v$ denotes tail-Gini functional.\\
 In this case,  $\hat{g}(u)=\frac{1}{(1-p)^{2}}\left[(1-p)u+4\tau\left(\frac{u^{2}}{2}-\frac{(1+p)u}{2}\right)+p^{2}-(1-2\tau)p\right]\mathbb{I}_{[p,1]}(u)$, $u\in[0,1]$, $\tau\in[0,\frac{1}{2}]$, and $g(1)=\hat{g}(1)=1$ in Theorem \ref{th.1}. Because $\hat{g}(u)$ is convex in $[0,1]$, $\hat{g}_{\ast}(u)=\hat{g}(u)$, and
 \begin{align*}
 \hat{g}_{\ast}'(u)=
 \begin{cases}
 \frac{1}{(1-p)^{2}}\left[(1-p)+4\tau\left(u-\frac{1+p}{2}\right)\right],~u\in[p,1],\\
 0,~u\in[0,p).
 \end{cases}
 \end{align*}
 Thus,
  \begin{align*}
   \sup_{X\in V (\mu,\sigma)}\mathrm{GS}_{p}^{\tau} (X)= \mu +\sigma\sqrt{\frac{3p+4\tau^{2}}{3(1-p)}},
\end{align*}
which can be obtained by the
worst-case distribution of r.v. $X_{\ast}$ characterized by
\begin{align*}
F_{X_{\ast}}^{-1}(u)=
\begin{cases}
\mu+\sigma\frac{\sqrt{3}\left[p(1-p)+4\tau\left(u-\frac{1+p}{2}\right)\right]}{(1-p)^{\frac{3}{2}}\sqrt{3p+4\tau^{2}}},~u\in[p,1],\\
\mu-\sigma\sqrt{\frac{3(1-p)}{3p+4\tau^{2}}},~u\in[0,p).
\end{cases}
\end{align*}
\end{example}

\begin{example}\label{ex.18}A random variable $X$, has its the extended Gini shortfall (EGS) with parameter $r>1$,
$p\in(0,1)$ and the loading
parameter $\tau \geq 0$,
  denoted by $\mathrm{EGS}_{p}^{\tau} (X)$ (see, e.g., Berkhouch et al., 2018), as
\begin{align*}
\mathrm{EGS}_{p}^{\tau} (X) = \mathrm{ES}_{p}(X) + \tau \mathrm{TEGini}_{p}(X),
\end{align*}
 where $\mathrm{TEGini}_{p}(X)$ denotes tail-based extended Gini coefficient (see Remark \ref{re.6}).\\
In this case, $\hat{g}(u)=\frac{1}{(1-p)^{2}}\left[(1-p)u+2\tau\left((1-u)^{r}+(1-p)^{r-1}u\right)+p(p-1)-2\tau(1-p)^{r-1}\right]\mathbb{I}_{[p,1]}(u)$, $u\in[0,1]$, $r>1$, $\tau\in[0,1/(2(r-1)(1-p)^{r-2})]$, and $g(1)=\hat{g}(1)=1$ in Theorem \ref{th.1}. Since $\hat{g}(u)$ is convex in $[0,1]$ (for instance, $r=\frac{3}{2},2,3$ with $\tau=0.3~\mathrm{and} ~p=0.2$, $\hat{g}(u)$ are plotted in Fig. 6(a)), $\hat{g}_{\ast}(u)=\hat{g}(u)$, and
 \begin{align*}
 \hat{g}_{\ast}'(u)=
 \begin{cases}
 \frac{1}{(1-p)^{2}}\left[(1-p)+2\tau\left(-r(1-u)^{r-1}+(1-p)^{r-1}\right)\right],~u\in[p,1],\\
 0,~u\in[0,p).
 \end{cases}
 \end{align*}
 Hence,
  \begin{align*}
   \sup_{X\in V (\mu,\sigma)}\mathrm{EGS}_{r,p}^{\tau} (X)= \mu +\sigma\sqrt{\frac{p}{1-p}+\frac{4\tau^{2}(1-p)^{2r-5}(r-1)^{2}}{2r-1}},
\end{align*}
which can be obtained by the
worst-case distribution of r.v. $X_{\ast}$ characterized by
\begin{align*}
F_{X_{\ast}}^{-1}(u)=
\begin{cases}
\mu+\sigma\frac{\sqrt{2r-1}\left[p(1-p)+2\tau\left(-r(1-u)^{r-1}+(1-p)^{r-1}\right)\right]}{(1-p)^{\frac{3}{2}}\sqrt{(2r-1)p+4\tau^{2}(1-p)^{2r-4}(r-1)^{2}}},~u\in[p,1],\\
\mu-\sigma\sqrt{\frac{(2r-1)(1-p)}{(2r-1)p+4\tau^{2}(1-p)^{2r-4}(r-1)^{2}}},~u\in[0,p).
\end{cases}
\end{align*}
\end{example}

\begin{example}\label{ex.19}  The shortfall of cumulative residual entropy of a random variable $X$, with $p\in(0,1)$ and the loading
parameter $\tau \geq 0$, denoted by $\mathrm{CRES}_{p}^{\tau} (X)$ (see Hu and Chen (2020)), given by
\begin{align*}
\mathrm{CRES}_{p}^{\tau} (X) = \mathrm{ES}_{p}(X) + \tau \mathrm{TCRE}_{p}(X),
\end{align*}
where $\mathrm{TCRE}_{p}(X)$ is the tail-based cumulative residual entropy (see Example \ref{ex.7}).\\
In this case,  $\hat{g}(u)=\left[\frac{u-p}{1-p}+\tau\frac{1-u}{1-p}\log\frac{1-u}{1-p}\right]\mathbb{I}_{[p,1]}(u)$, $u\in[0,1]$, $\tau\in[0,1]$, and $g(1)=\hat{g}(1)=1$ in Theorem \ref{th.1}. Because $\hat{g}(u)$ is convex in $[0,1]$, $\hat{g}_{\ast}(u)=\hat{g}(u)$, and
 \begin{align*}
 \hat{g}_{\ast}'(u)=
 \begin{cases}
  \frac{1-\tau}{1-p}-\frac{\tau}{1-p}\log\frac{1-u}{1-p},~u\in[p,1],\\
 0,~u\in[0,p).
 \end{cases}
 \end{align*}
 Thus,
  \begin{align*}
   \sup_{X\in V (\mu,\sigma)}\mathrm{CRES}_{p}^{\tau} (X)= \mu +\sigma\sqrt{\frac{p+\tau^{2}}{1-p}},
\end{align*}
which can be obtained by the
worst-case distribution of r.v. $X_{\ast}$ characterized by
\begin{align*}
F_{X_{\ast}}^{-1}(u)=
\begin{cases}
\mu+\sigma\frac{(p-\tau)-\tau\log\frac{1-u}{1-p}}{\sqrt{(1-p)(p+\tau^{2})}},~u\in[p,1],\\
\mu-\sigma\sqrt{\frac{1-p}{p+\tau^{2}}},~u\in[0,p).
\end{cases}
\end{align*}
\end{example}

\begin{example}\label{ex.20} A random variable $X$, has its the shortfall of cumulative residual Tsallis entropy with order $\alpha>0$, $\alpha\neq1$,
$p\in(0,1)$ and the loading
parameter $\tau \geq 0$,
  denoted by $\mathrm{CRTES}^{\tau}
_{\alpha,p}(X)$ (see, e.g., Zuo and Yin (2023)), as
\begin{align*}
\mathrm{CRTES}^{\tau}
_{\alpha,p}(X)= \mathrm{ES}_{p}(X) + \tau \mathrm{TCRTE}_{\alpha,p}(X),
\end{align*}
where $\mathrm{TCRTE}_{\alpha,p}(X)$ is the tail-based cumulative residual Tsallis entropy of order $\alpha$ (see Eq. (\ref{a12})).\\
In this case, $\hat{g}(u)=\left[\frac{u-p}{1-p}-\frac{\tau}{\alpha-1}\left(\frac{1-u}{1-p}-\left(\frac{1-u}{1-p}\right)^{\alpha}\right)\right]\mathbb{I}_{[p,1]}(u)$, $u\in[0,1]$, $\alpha>0$, $\alpha\neq1$, $\tau\in[0,1]$, and $g(1)=\hat{g}(1)=1$ in Theorem \ref{th.1}. Since $\hat{g}(u)$ is convex in $[0,1]$ (for instance, $r=\frac{2}{3},\rightarrow1,=2$ with $\tau=0.5~\mathrm{and} ~p=0.2$, $\hat{g}(u)$ are plotted in Fig. 6(b)), $\hat{g}_{\ast}(u)=\hat{g}(u)$, and
 \begin{align*}
 \hat{g}_{\ast}'(u)=
 \begin{cases}
  \frac{1}{1-p}-\frac{\tau}{\alpha-1}\left[\frac{-1}{1-p}+\frac{\alpha}{1-p}\left(\frac{1-u}{1-p}\right)^{\alpha-1}\right],~u\in[p,1],\\
 0,~u\in[0,p).
 \end{cases}
 \end{align*}
 Then, we have
  \begin{align*}
   \sup_{X\in V (\mu,\sigma)}\mathrm{CRTES}^{\tau}_{\alpha,p}(X)= \mu +\sigma\sqrt{\frac{(2\alpha-1)p+\tau^{2}}{(2\alpha-1)(1-p)}},~\alpha>\frac{1}{2},
\end{align*}
which can be obtained by the
worst-case distribution of r.v. $X_{\ast}$ characterized by
\begin{align*}
F_{X_{\ast}}^{-1}(u)=
\begin{cases}
\mu+\sigma\frac{\sqrt{2\alpha-1}\left[p+\frac{\tau}{\alpha-1}\left(1-\alpha\left(\frac{1-u}{1-p}\right)^{\alpha-1}\right)\right]}{\sqrt{(1-p)[(2\alpha-1)p+\tau^{2}]}},~u\in[p,1],\\
\mu-\sigma\sqrt{\frac{(2\alpha-1)(1-p)}{(2\alpha-1)p+\tau^{2}}},~u\in[0,p).
\end{cases}
\end{align*}
\end{example}
\begin{remark}
Note that let $\tau=0$ in Examples \ref{ex.17}-\ref{ex.20}, above shortfalls will reduce to $\mathrm{ES}_{p}(X)$:
\begin{align*}
\sup_{X\in V (\mu,\sigma)}\mathrm{ES}_{p}(X)=\mu+\sigma\sqrt{\frac{p}{1-p}},
\end{align*}
which can be obtained by the
worst-case distribution of r.v. $X_{\ast}$ characterized by
\begin{align*}
F_{X_{\ast}}^{-1+}(u)=
\begin{cases}
\mu+\sigma\sqrt{\frac{p}{1-p}},~u\in[p,1],\\
\mu-\sigma\sqrt{\frac{1-p}{p}},~u\in[0,p).
\end{cases}
\end{align*}
This coincides with (5.1) of Zhao et al. (2024).
\end{remark}

\begin{remark}
From Proposition \ref{pro.2}, let $g(u)=-2(-u+u^{2}),$ $-2(1-p)^{r-2}(u^{r}-u),~r>1$, $-u\log(u)$ and $\frac{1}{\alpha-1}[u-u^{\alpha}], ~\alpha>0,~\alpha\neq1$, $u\in[0,1]$, respectively, we also obtain worst-cases of Gini shortfall ($\mathrm{GS}_{p}^{\tau} (X)$), Extended Gini shortfall ($\mathrm{EGS}_{p}^{\tau} (X)$), shortfall of cumulative residual entropy ($\mathrm{CRES}_{p}^{\tau} (X) $) and shortfall of cumulative residual Tsallis entropy with order $\alpha$ ($\mathrm{CRTES}^{\tau}
_{\alpha,p}(X)$) as in Examples \ref{ex.17}-\ref{ex.20}.

In addition, we find that for same $p$ and $\tau$, $\sup_{X\in V (\mu,\sigma)}\mathrm{GS}_{p}^{\tau} (X)>\sup_{X\in V (\mu,\sigma)}\mathrm{CRES}_{p}^{\tau} (X)$, $\sup_{X\in V (\mu,\sigma)}\mathrm{EGS}_{r=2,p}^{\tau} (X)=\sup_{X\in V (\mu,\sigma)}\mathrm{GS}_{p}^{\tau} (X) $, $\lim_{\alpha\rightarrow1}\sup_{X\in V (\mu,\sigma)}\mathrm{CRTES}^{\tau}_{\alpha,p}(X)=\sup_{X\in V (\mu,\sigma)}\mathrm{CRES}_{p}^{\tau} (X)$.
\end{remark}
 \section{Numerical illustration\label{sec:7}}
In this section, we consider daily stock returns of three stock companies from the Nasdaq stock market (Cisco Sys. Inc.
(CSCO), Apple Inc. (AAPL),  eBay Inc. (EBAY)) for the period  2023-04-25 to 2024-04-24, (For the data, see http://www.nasdaq.com/), and denote by $X_{1}$, $X_{2}$ and $X_{3}$, respectively. We can compute their mean and variance respectively written as $(\mu=0.04371627,\sigma^{2}=0.191021554)$, $(\mu=0.123873016
,\sigma^{2}=3.204667195)$ and $(\mu=0.021860317,\sigma^{2}=0.39813437)$.

Now, we compute their sharp upper bounds for premium principles. Let $\alpha=2/3~\mathrm{and} ~3$, $r=3/2~\mathrm{and}~3$ in Section \ref{sec:6.1}.  The sharp upper bounds of different premium principles of the same $X_{i},~i=1,2,3$ for $\kappa\in[0,1]$ are shown in Figs. 7 (a)-(c), respectively. The sharp upper bounds of same premium principles, including $\mathrm{BGini}$, $\mathrm{B}\mathcal{CE}$, $\mathrm{BCT}_{\frac{2}{3}}$, $\mathrm{BCT}_{3}$, $\mathrm{BEGini}_{\frac{3}{2}}$ and $\mathrm{BEGini}_{3}$,  of $X_{1}$, $X_{2}$ and $X_{3}$ for $\kappa\in[0,1]$ are presented in Figs. 8 (a)-(f), respectively.

In Fig. 7 (a), we plot sharp upper bounds of different premium principles of $X_{1}$ for $\kappa\in[0, 1]$ by Matlab software.  It is seen from the figure that  sharp upper bounds of all premium principles are (linear) increasing in $\kappa$. For fixed $\kappa$, $\mathrm{BEGini}_{3}(X_{1})$ is largest in all premium principles, and $\mathrm{BCT}_{3}(X_{1})$ is least in all premium principles. Moreover, in Figs. 7 (b) and (c), we plot sharp upper bounds of different premium principles of $X_{2}$ and $X_{3}$ for $\kappa\in [0, 1]$ by Matlab software, respectively. The results of Figs. 7 (b) and (c) are similar to those of Fig. 7 (a).

Figs. 8 (a)-(f) plot sharp upper bounds of $\mathrm{BGini}$, $\mathrm{B}\mathcal{CE}$, $\mathrm{BCT}_{\frac{2}{3}}$, $\mathrm{BCT}_{3}$, $\mathrm{BEGini}_{\frac{3}{2}}$ and $\mathrm{BEGini}_{3}$  of $X_{1}$, $X_{2}$ and $X_{3}$ for $\kappa\in [0, 1]$ by Matlab software.  As we see in figures, for fixed $\kappa$, the sharp upper bounds of premium principles of $X_{2}$ are largest in three companies ($X_{i},~i=1,2,3$), and the sharp upper bounds of premium principles of $X_{1}$ are almost least in three companies ($X_{i},~i=1,2,3$).

Next, we calculate their sharp upper bounds of shortfalls.
Let $\tau=0.5$, $\alpha=2/3~\mathrm{and} ~3$, $r=3$ in Section \ref{sec:6.2} (or Examples \ref{ex.17}-\ref{ex.20}). The sharp upper bounds of different shortfalls of the same $X_{i},~i=1,2,3$ for $p\in[0.9,1)$ are shown in Figs. 9 (a)-(c), respectively. The sharp upper bounds of same shortfalls, including $\mathrm{GS}_{p}$, $\mathrm{CRES}_{p}$, $\mathrm{CRTES}_{\frac{2}{3},p}$, $\mathrm{CRTES}_{3,p}$ and $\mathrm{EGS}_{3,p}$,  of $X_{1}$, $X_{2}$ and $X_{3}$ for $p\in[0.9,1)$ are presented in Figs. 10 (a)-(f), respectively.

In Fig. 9 (a), we plot sharp upper bounds of different shortfalls of $X_{1}$ for $p\in [0.9, 1)$ by Matlab software.  From the figure we observe that  sharp upper bounds of all shortfalls are increasing in $p$. For fixed $p$, $\mathrm{CRTES}_{\frac{2}{3},p}(X_{1})$ is largest in all shortfalls, and $\mathrm{EGS}_{3,p}(X_{1})$ is least in all shortfalls. Furthermore, in Figs. 9 (b) and (c), we plot sharp upper bounds of different shortfalls of $X_{2}$ and $X_{3}$ for $p\in [0.9, 1)$ by Matlab software, respectively. The results of Figs. 9 (b) and (c) are similar to those of Fig. 9 (a).

Figs. 10 (a)-(e) plot sharp upper bounds of $\mathrm{GS}_{p}$, $\mathrm{EGS}_{3,p}$, $\mathrm{CRTES}_{\frac{2}{3},p}$, $\mathrm{CRES}_{p}$ and $\mathrm{CRTES}_{3,p}$,  and of $X_{1}$, $X_{2}$ and $X_{3}$ for $p\in [0.9, 1)$ by Matlab software.  As we see in figures, for fixed $p$, the sharp upper bounds of shortfalls of $X_{2}$ are largest in three companies ($X_{i},~i=1,2,3$), and the sharp upper bounds of shortfalls of $X_{1}$ are almost least in three companies ($X_{i},~i=1,2,3$).
 \section{Concluding remarks\label{sec:8}}
 This paper has considered the worst-case of weighted entropy for general distributions when only partial information (mean and variance) is known, and also has focused on the worst-case of distortion riskmetrics for general distributions when only partial information is available. These results are the extensions of results of Zhao et al. (2024) and Shao and Zhang (2023a, 2023b). Furthermore, Zhao et al. (2024) also has derived extremal cases of general class of distortion risk measures for symmetric, unimodal and unimodal-symmetric random variables. We will study the worst-cases of weighted entropy and distortion riskmetrics for symmetric, unimodal and unimodal-symmetric  distributions, hoping these results could be reported  in the future.

\section*{ CRediT authorship contribution statement}
\noindent$\mathbf{Baishuai~ Zuo:}$ Investigation, Methodology, Writing-original draft, Writing-review \& editing, Software.\\
$\mathbf{ Chuancun~ Yin:}$ Conceptualization, Investigation, Methodology, Supervision, Validation, Writing-original draft, Writing-review \& editing.
\section*{Acknowledgments}

\noindent  The research was supported by the National Natural Science Foundation of China (No. 12071251, 12371472)
\section*{Declaration of competing interest}
\noindent There is no competing interest.

\medskip

\end{document}